# A Large Sensor Foundation Model Pretrained on Continuous Glucose Monitor Data for Diabetes Management

Junjie Luo, M.S.,[1,2] Abhimanyu Kumbara, M.S.,[3] Mansur Shomali, M.D.,[3] Rui Han, M.S.,[2] Anand Iyer, Ph.D., M.B.A.,[3] Grazia Aleppo, M.D.,[4] Ritu Agarwal, Ph.D., M.B.A.,[2] Gordon Gao, Ph.D., M.B.A.,[2] *

Author affiliations
[1] School of Medicine, Johns Hopkins University
[2] The Center for Digital Health and Artificial Intelligence, Johns Hopkins University
[3] Welldoc Inc.
[4] Feinberg School of Medicine, Northwestern University

* Corresponding author: Prof. Gordon Gao, Johns Hopkins University, co-director of the Center for Digital Health and AI (CDHAI), Email: gordon.gao@jhu.edu., ORCID: 0000-0002-2336-9682

# Abstract

Continuous glucose monitoring (CGM) combined with AI offers new opportunities for proactive diabetes management through real-time glucose forecasting. However, most existing models are task-specific and lack generalization across patient populations. Inspired by the autoregressive paradigm of large language models, we introduce CGM-LSM, a Transformer decoder-based Large Sensor Model (LSM) pretrained on 1.6 million CGM records from patients with different diabetes types, ages, and genders. We model patients as sequences of glucose time steps to learn latent knowledge embedded in CGM data and apply it to the prediction of glucose readings for a 2-hour horizon. Compared with prior methods, CGM-LSM significantly improves prediction accuracy and robustness: a 48.51% reduction in root mean square error in one-hour horizon forecasting and consistent zero-shot prediction performance across held-out patient groups. We analyze model performance variations across patient subgroups and prediction scenarios and outline key opportunities and challenges for advancing CGM foundation models.

# Introduction

Diabetes imposes significant burdens on individuals, families, and healthcare providers. Uncontrolled diabetes has been associated with a variety of severe medical complications, including heart attacks[1], kidney failure[2], and diabetic neuropathy[3]. In 2021, 38.4 million Americans (11.6% of the U.S. population) had diabetes, making it the eighth leading cause of death[4]. By 2022, diabetes management costs had soared to approximately $412.9 billion annually (25% of total healthcare expenditures)[4]. The rising societal and economic burden of diabetes combined with a growing global shortage of healthcare professionals necessitates new approaches to diabetes care and prevention[5].

Effective self-management is crucial for people with diabetes to manage their condition and avoid complications. To inform their daily decision-making processes, patients need actionable information. Artificial intelligence (AI) holds considerable promise to provide such information by generating personalized interventions tailored to individual needs. However, most AI applications in diabetes management have been centered on predicting the overall risk of diabetes progression over long time horizons, and such predictions are difficult for individuals to convert into daily diabetes management activities. This shortcoming underscores the need for AI solutions in diabetes management that provide timely, micro-level, personalized, actionable predictions to enable patients to improve their self-care[6–8].

In recent years, continuous glucose monitoring (CGM) systems, which continuously measure glucose values every five minutes, have emerged as a critical tool in diabetes control[9,10]. One particularly promising approach to meeting the patient needs described above lies in combining CGM with AI to predict near-future glucose values[11]. However, little research has addressed this task, which is inherently challenging due to the complexity and variability of individual behaviors captured by CGM. This makes it difficult to accurately predict a patient's glucose levels for the next two hours[12]. Studies using simulated CGM data or datasets from a limited number of patients report poor prediction performance[13–18]. Such poor performance can be caused not only by inadequate data but also by training the model using random initialization with zero glucose generation knowledge. In this context, the question of how to utilize the hidden knowledge within CGM data patterns to empower prediction models remains unaddressed.

To address this gap, we present a new approach to glucose prediction for diabetes management that harnesses the pretraining technique exemplified in large language models (LLMs). We pretrained the model to learn the latent glucose generation mechanisms hidden in massive CGM data. Considering that language is represented as a sequence of tokens and LLMs learn through next-token prediction, we also modeled person-generated glucose data as a sequence of time steps (e.g., every five minutes) and proposed a large sensor model (LSM) to learn massive sensor data with next-step prediction. We hypothesize that an LSM pretrained on CGM data (CGM-LSM) could learn glucose fluctuation patterns and achieve high performance in predicting glucose values.

We developed the CGM-LSM on a large CGM dataset with 15.96 million glucose records from 592 patients with diabetes. We validated its effectiveness in improving near-future glucose prediction using metrics of accuracy and robustness. Benchmarked against the public OhioT1DM dataset, CGM-LSM achieved a root Mean Squared Error (rMSE) of 15.90 mg/dL for one-hour horizon prediction, which is 48.51% lower than that of the baseline approaches, even though our model was not trained on the OhioT1DM dataset[16]. The model's robustness was confirmed through analyses assessing generalizability across varying prediction horizons, performance metrics, diabetes types, unseen patients, and age and gender groups. Notably, prediction accuracy declined during high-variability contexts, such as daytime hours, mealtimes, and glucose extremes. This highlights the need for future studies to address underrepresented scenarios.

This study proposes the LSM pretrained with glucose sensor data and demonstrates its significant improvement in near-future glucose prediction. It offers a new methodology not only for timely and personalized diabetes self-management but also for learning from massive sensor data.

# Results

## Datasets

The key data point in our study is an "instance," defined as a combination of a patient and an observation datetime. This datetime represents the time when the model is triggered to observe the patient's records and generate predictions. As shown in Fig. 1B, inclusion as a valid instance requires 288 CGM entries from the 24 hours prior to the datetime and 24 entries in the subsequent 2 hours, for a total of 26 hours of CGM data.

| | Records (Patients) | Type 1 Diabetes (T1D) | | | | | | Type 2 Diabetes (T2D) | | | | | |
|---|---|---|---|---|---|---|---|---|---|---|---|---|---|
| | | **Complete Dataset** | Dataset | Training | Internal | Temporal | Held-Out | **Complete Dataset** | Dataset | Training | Internal | Temporal | Held-Out |
| **Patients** | 15,961,183 (592) | 7,788,836 (291) | 779,111 (290) | 560,184 (257) | 61,979 (256) | 68,762 (257) | 88,186 (33) | 8,172,347 (301) | 817,125 (301) | 597,975 274) | 66,257 (273) | 73,293 (273) | 79,600 (27) |
| **Age Group** | | | | | | | | | | | | | |
| **18-39 years old** | 3,425,955 (129) | 2,901,601 (109) | 290,615 (108) | 209,054 (94) | 23,321 (94) | 25,837 (94) | 32,403 (14) | 524,354 (20) | 52,077 (20) | 39,252 (19) | 4,440 (19) | 4908 (19) | 3,477 (1) |
| **40-64 years old** | 8,040,123 (299) | 3,249,786 (123) | 324,649 (123) | 231,576 (108) | 25,408 (107) | 28,412 (108) | 39,253 (15) | 4,790,337 (176) | 478,441 (176) | 347,663 (158) | 38,481 (158) | 42,483 (158) | 49,814 (18) |
| **65 + years old** | 4,495,105 (164) | 1,637,449 (59) | 163,847 (59) | 119,554 (55) | 13,250 (55) | 14,513 (55) | 16,530 (4) | 2,857,656 (105) | 286,607 (105) | 211060 (97) | 23,336 (96) | 25902 (96) | 26,309 (8) |
| **Gender Group** | | | | | | | | | | | | | |
| **Female** | 7,133,594 (283) | 4,150,047 (165) | 415,869 (164) | 295,753 (144) | 32,910 (144) | 36,478 (144) | 50,728 (20) | 2,983,547 (118) | 297,557 (118) | 224,985 (109) | 24,768 (108) | 27,582 (108) | 20,222 (9) |
| **Male** | 8,827,589 (309) | 3,638,789 (126) | 363,242 (126) | 264,431 (113) | 29,069 (112) | 32,284 (113) | 37,458 (13) | 5,188,800 (183) | 519,568 (183) | 372,990 (165) | 41,489 (165) | 45,711 (165) | 59,378 (18) |

**Table 1.** Data description for the Welldoc dataset, showing the number of instances and patient counts across subsets by diabetes type (Type 1 Diabetes [T1D] and Type 2 Diabetes [T2D]), age groups, and gender. Subsets include the complete dataset, training, internal, temporal, and held-out sets.

To pretrain our CGM-LSM model, we constructed a dataset of 592 individuals diagnosed with diabetes (291 T1D and 301 T2D). The dataset contains 15,961,183 valid instances, 7,788,836 from T1D patients and 8,172,347 from T2D. The distribution is consistent across various demographics, with 2,901,601 instances involving T1D individuals aged 18-39 and 524,354 involving their T2D counterparts. Female T1D patients account for 4,150,047 instances, while T2D females contribute 2,983,547 instances.

To benchmark its performance with baseline methods, we evaluate our pretrained model with the OhioT1DM dataset[19]. After a similar preprocessing method, we get 12 T1D patients with a total of 58,414 instances (44,275 from train, 4,925 from validation, and 9,214 for test).

# CGM-LSM Pretraining and Generation

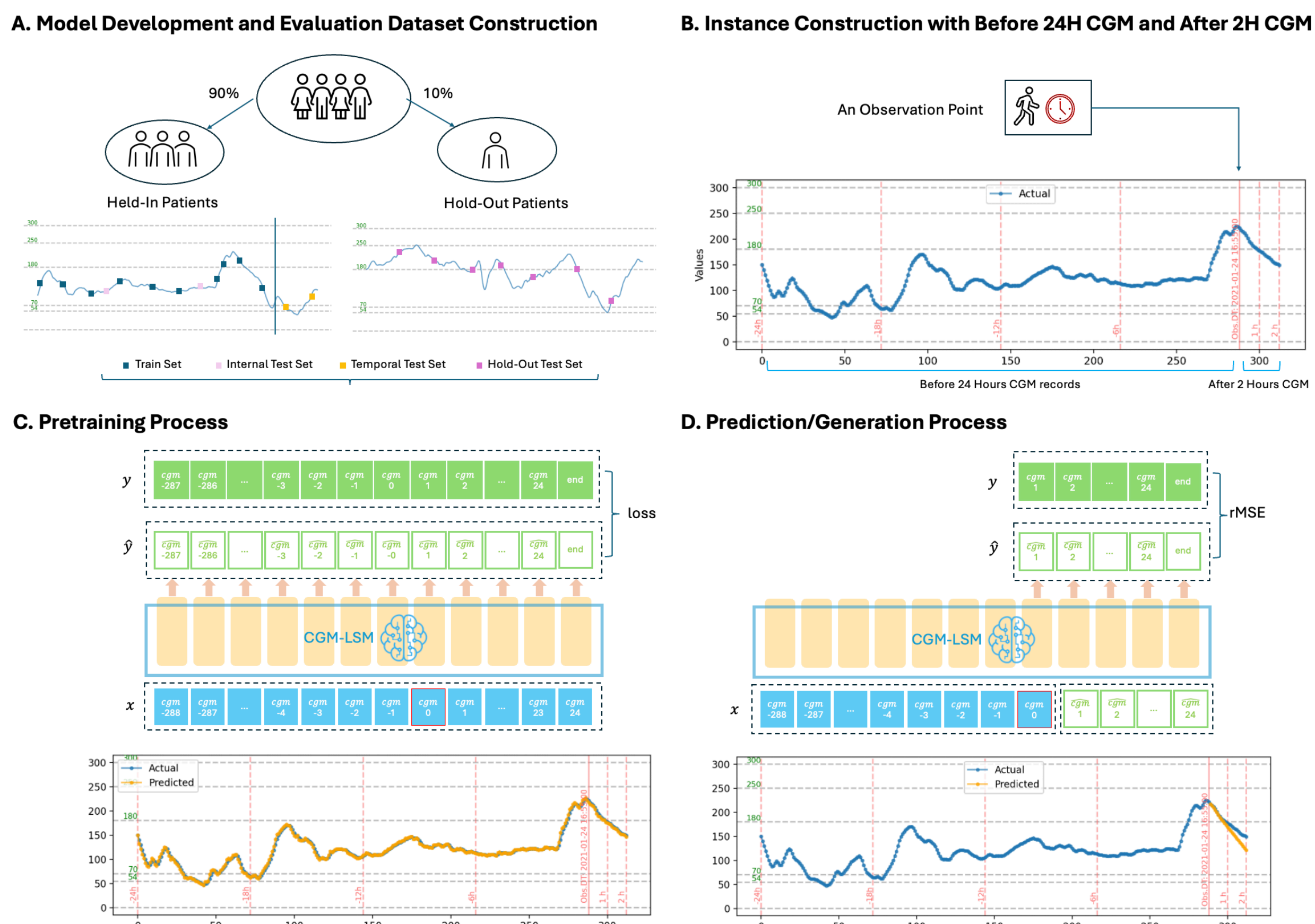

**Figure 1. The workflow of dataset construction and CGM-LSM development**. **A**. The instances selection process to construct internal test set, temporal test set, held-out test set, and training set. **B**. For one instance, the input-output pair construction process. Each instance is a combination of a patient and an instance datetime. The instance contains 288 CGM records before 24 hours and 24 CGM records after 2 hours. **C**. The pretraining process for one instance given 26-hour CGM records. **D.** The prediction (generation) process for one instance with 24-hour CGM records.

As shown in Fig. 1A, to test the model's robustness and generalizability, we randomly split the dataset into two patient groups: held-in and held-out. Data from held-out patients (10% of patients) were excluded from the training phase and reserved exclusively as the "held-out" test set. Each "held-in" patient's instances were chronologically sorted based on observation datetimes. The latter 10% were reserved to form the "temporal" test set, intended for evaluating model performance on unseen future periods. The remaining 90% of instances were further randomly split, with 10% used as the "internal" test set and 80% as the "training" dataset. This stratification allowed the model to be evaluated on unseen future periods and on unseen patients. To mitigate overfitting and enhance computational efficiency, a random 10% down-sampling was applied to each dataset. The final numbers of patients and instances are summarized in Table 1.

The training process (Fig. 1C) employed transformer-decoder architecture[20] to develop CGM-LSM through autoregression training. This design is particularly advantageous for learning temporal dependencies in continuous glucose data. Compared to convolutional neural networks (CNNs)[21], which are limited in receptive field unless heavily stacked, and recurrent models, such as RNNs[22], LSTMs[23], and GRUs[24], which suffer from vanishing gradients and sequential bottlenecks, the Transformer decoder enables efficient access to all past timesteps via self-attention. This allows the model to flexibly capture both short- and long-range glucose dynamics without relying on fixed context windows or handcrafted lag features. Furthermore, unlike Transformer encoder-only architectures, which process sequences non-autoregressively and are commonly used for classification or embedding tasks, the decoder structure is well-suited for sequence generation. It operates autoregressively, predicting each future glucose value based on previously generated values, making it highly aligned with the objective of pretraining via next-step prediction. This autoregressive setup closely mirrors the downstream forecasting task and enables the model to learn structured temporal patterns in a self-supervised manner. As a result, CGM-LSM demonstrates improved generalization and robustness across patients and time horizons.

In the training process, the model predicts the glucose value in the next time step by iteratively processing each time step across a 26-hour window. To model glucose trajectories, we discretize glucose values into categorical tokens and apply a cross-entropy loss over the token vocabulary. This design enables us to utilize autoregressive pretraining frameworks used in large language models. This is because glucose values have nonlinear clinical significance: a change from 80 to 100 mg/dL (crossing from hypoglycemia to euglycemia) carries different clinical implications than a change from 180 to 200 mg/dL, which remains within a hyperglycemic range. By learning token embeddings rather than using raw values, the model can capture these asymmetric and context-dependent relationships effectively. This token-based formulation also aligns well with recent time-series foundation models such as Time-LLM[25] and Chronos[26], which demonstrate that discretization combined with learned embeddings can generalize well across complex temporal sequences. In the generation process (Fig. 1D), for each instance, the model takes the previous 24 hours of CGM data as input to predict the glucose values for the subsequent two hours.

## Evaluation of Pretraining Effectiveness

To assess whether the model benefits from large-scale pretraining, we begin by examining what it learns about the structure of glucose values during pretraining. The first question is whether the model, despite using a categorical tokenization scheme, can recover the inherent continuity and ordering of numerical glucose values. We tokenize glucose values into discrete bins to enable autoregressive sequence modeling with a Transformer decoder. While this discretization simplifies model input and increases semantic potentials, it may risk discarding the continuous nature of glucose signals unless the model can infer this structure implicitly.

To evaluate this, we visualize the CGM-LSM's token embeddings after pretraining using t-SNE (Figure 2). Each point corresponds to an individual glucose token, and points are colored by their glucose value range. The embeddings form smooth, coherent trajectories in the low-dimensional space, with adjacent tokens aligning in a physiologically meaningful order. This spatial organization suggests that the model learns to encode ordinal and contextual relationships among glucose values through pretraining. These results indicate that categorical tokenization, combined with large-scale pretraining, enables the model to learn the latent numerical structure of CGM data.

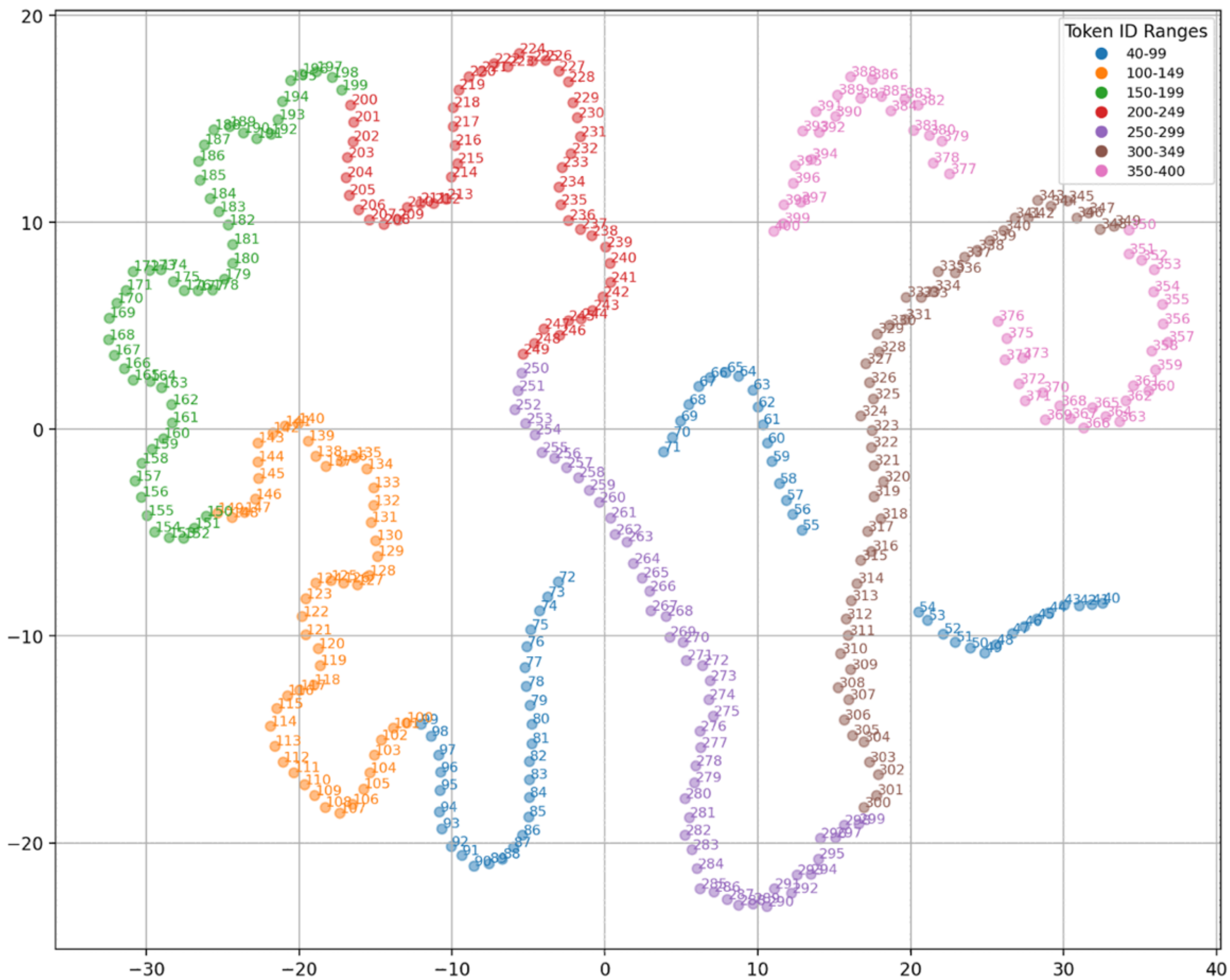

**Figure 2. CGM-LSM Token Embeddings (Colored by Glucose Range).** Each point represents a token ID, which directly corresponds to a glucose value. Colors indicate glucose ranges commonly used in clinical interpretation.

To evaluate whether the model's learned token representations and attention-based dynamics contribute to meaningful improvements in forecasting, we assess CGM-LSM's performance on the publicly available OhioT1DM dataset[19]. This benchmark was not used during pretraining or fine-tuning, providing a rigorous test of zero-shot generalization. As shown in Table 2, CGM-LSM significantly outperforms traditional sequence models across all forecasting horizons. At the 30-minute horizon, it achieves an rMSE of 9.02 mg/dL, representing more than a 65% reduction in error compared to baseline models. At the 1-hour horizon, CGM-LSM maintains strong performance with an rMSE of 15.90 mg/dL, representing a 48.51% reduction compared to vanilla Transformer[27]. At the 2-hour prediction horizon, CGM-LSM achieves an rMSE of 26.88 mg/dL, outperforming the strongest baselines by 26.67%. These results indicate that CGM-LSM generalizes well across patients and longer time horizons.

We believe CGM-LSM's strong performance on OhioT1DM can be attributed to the pretraining on a large and diverse corpus of CGM sequences spanning different patient types, diabetes subgroups, and glycemic patterns. Through exposure to a broad distribution of glucose trajectories, the model learns to capture latent structure and hidden physiological patterns that small datasets alone cannot reveal. This enables CGM-LSM to make accurate forecasts even for unseen patients by utilizing its attention mechanism and token representations. In contrast, models trained directly on limited datasets like OhioT1DM are prone to underfitting and often fail to exploit the full modeling capacity of deep architectures.

| **Model** | **Dataset** | | **rMSE-30m** | **rMSE-1h** | **rMSE-2h** |
|---|---|---|---|---|---|
| LSTM | OhioT1DM | | 36.022 (0.551) | 37.17 (0.513) | 38.703 (0.461) |
| RNN | | | 36.102 (0.552) | 37.344 (0.512) | 38.952 (0.458) |
| GRU | | | 37.555 (0.554) | 38.573 (0.517) | 40.108 (0.463) |
| Transformer | | | 27.886 (0.463) | 30.869 (0.462) | 36.653 (0.47) |
| Informer | | | 35.197 (0.501) | 36.962 (0.485) | 40.204 (0.463) |
| Autoformer | | | 36.08 (0.552) | 38.352 (0.542) | 41.395 (0.515) |
| CGMLSM | | | 9.024 (0.168) | 15.895 (0.283) | 26.876 (0.43) |
| CGM-LSM | WellDoc T1D | Internal Test | 8.403 (0.066) | 16.049 (0.118) | 28.277 (0.188) |
| | | Temporal Test | 9.155 (0.068) | 17.013 (0.118) | 29.426 (0.184) |
| | | Held-Out Test | 8.926 (0.056) | 16.905 (0.101) | 29.812 (0.16) |
| | WellDoc T2D | Internal Test | 7.441 (0.055) | 13.418 (0.094) | 22.649 (0.147) |
| | | Temporal Test | 8.025 (0.058) | 14.073 (0.095) | 23.216 (0.143) |
| | | Held-Out Test | 7.772 (0.055) | 13.877 (0.091) | 23.494 (0.143) |

**Table 2:** Comparative performance of predictive models for future glucose levels, using Root Mean Square Errors (rMSE). Each entry displays the mean rMSE followed by the confidence interval width in parentheses, indicating the range within which the true mean is expected to lie with 95% confidence.

## CGM-LSM Performance for Temporal Test and Held-out Patients

Assessing the model's ability to generalize to real-world scenarios, such as future periods and unseen patients, is crucial for practical disease management. To this end, with the advantage of the Welldoc dataset, we focus on two key evaluation settings: the **temporal test**, which simulates future glucose prediction for existing patients based on their prior CGM history, and the **held-out test**, which evaluates zero-shot performance on entirely new patients at onboarding. We hypothesized that both settings would pose greater challenges for CGM-LSM than the internal test set, due to temporal shifts in the future set and patient variability in the held-out set—both of which deviate from the training data distribution.

Despite these added complexities, CGM-LSM's performance gap between the temporal and held-out sets versus the internal set remained minimal. For T1D-2H predictions, rMSE was 28.28 mg/dL for the internal set, increasing slightly to 29.43 mg/dL (+4.07%) for the temporal set and 29.81 mg/dL (+5.43%) for the held-out set. Similarly, for T2D-2H predictions, rMSE increased modestly from 22.65 mg/dL for the internal set to 23.22 mg/dL (+2.51%) for the temporal set and 23.49 mg/dL (+3.73%) for the held-out set. Though these results show that while temporal and held-out predictions posed additional challenges, CGM-LSM maintained high accuracy in both settings. Notably, performance on the temporal test was consistently better than on the held-out set, suggesting that access to a patient's prior CGM data confers an advantage. This indicates that continued pretraining or adaptation using a patient's longitudinal data may further enhance personalized forecasting accuracy.

## Performance Across Diabetes Type, Age, and Gender

Diabetes is a complicated disease with distinctive glucose generation mechanisms among patients. A patient's diabetes type, age, and gender significantly influence how glucose levels evolve over time[28,29]. Considering this heterogeneity, we analyzed CGM-LSM's performance across diabetes type, age, and gender.

CGM-LSM demonstrated strong performance for both T1D and T2D patients. For T1D patients, rMSE scores were 8.93mg/dL (30min), 16.91mg/dL (1H), and 29.81mg/dL (2H). Performance was better for T2D patients, with lower rMSE scores of 7.77mg/dL (30min), 13.88mg/dL (1H), and 23.49mg/dL (2H). These results underscore the model's robustness across diabetes types. For T2D, the rMSE rose slightly by 3.71% from 22.65mg/dL (internal) to 23.49mg/dL (held-out), as shown in Table 2. For T1D, the rMSE increase was larger, rising by 5.41% from 28.28mg/dL (internal) to 29.81mg/dL (held-out), which was both proportionally and absolutely higher. This indicates that CGM-LSM has higher generalizability for unseen T2D patients than T1D patients.

Figure 2.a shows rMSE-2H performance differences across age groups and evaluation sets. These results indicate stable CGM-LSM performance across age groups. In T1D held-out sets, rMSE scores ranged from 24.77 to 31.78mg/dL, or 83.09% to 106.71% of the overall rMSE (29.81mg/dL). For T1D, rMSEs peaked in the 18-39 group, decreased in the 40-64 group, and stabilized further in the 65+ cohort. The 65+ cohort also had the smallest gap between internal and held-out sets. These findings suggest that our model provides greater accuracy and robustness for older groups, possibly due to differences in disease management or progression. For T2D, CGM-LSM showed a high error rate in the held-out set for the 18-39 group. This spike likely reflects the small sample size, with only one T2D patient in the held-out set (Table 1). Excluding this outlier, T2D rMSEs were lower and more uniform across age groups compared to T1D. This indicates greater accuracy and robustness for T2D.

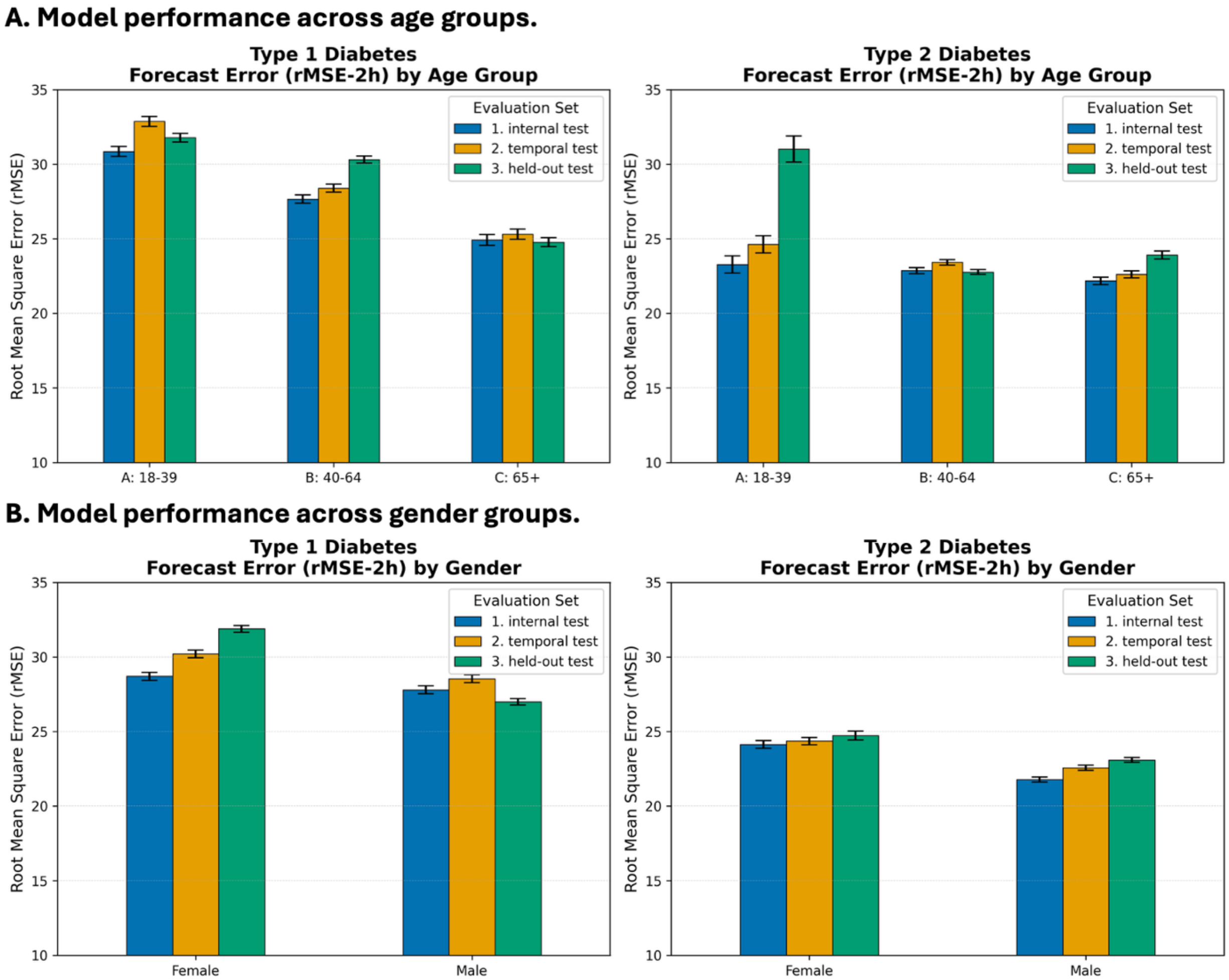

**Figure 3: Model Performance in 2-Hour Glucose Prediction Across Age and Gender. A.** Model performance across age groups (18-39, 40-64, 65+), showing rMSE-2H values for Type 1 and Type 2 Diabetes prediction across different test sets. **B.** Model performance across gender, showing rMSE-2H values for Type 1 and Type 2 Diabetes prediction across different test sets.

Figure 2.B highlights CGM-LSM's performance across gender, showing stable results across groups. For T1D held-out sets, rMSE scores ranged from 27.00 to 31.88mg/dL (90.57%–106.94% of the overall 29.81mg/dL). Males consistently had better prediction performance than females across all evaluation sets for both T1D and T2D. Among T1D patients, the rMSE gap between internal and held-out sets was notably larger for females (3.19mg/dL) than for males (-0.8mg/dL). These findings suggest the need for tailored model improvements to better accommodate unseen female T1D patients.

| Group | Std Glucose (mg/dL) | Coefficient of Variation | Mean Rate of Change (mg/dL/5min) | TIR (70–180 mg/dL) (%) | TAR (>180 mg/dL) (%) |
|---|---|---|---|---|---|
| **Type 1 Diabetes** | 48.39 (18.89) | 0.2986 (0.0931) | 3.57 (1.23) | 0.6536 (0.2284) | 0.3200 (0.2336) |
| **Type 2 Diabetes** | 37.68 (16.59) | 0.2309 (0.0812) | 3.08 (1.27) | 0.6940 (0.2701) | 0.2985 (0.2707) |
| **Age 18–39** | 49.94 (20.39) | 0.2934 (0.0954) | 3.88 (1.38) | 0.6143 (0.2544) | 0.3648 (0.2588) |
| **Age 40–64** | 41.49 (18.44) | 0.2567 (0.0969) | 3.26 (1.18) | 0.6770 (0.2602) | 0.3043 (0.2626) |
| **Age 65+** | 40.59 (15.82) | 0.2579 (0.0812) | 3.00 (1.21) | 0.7143 (0.2194) | 0.2748 (0.2208) |
| **Female** | 45.18 (19.22) | 0.2742 (0.0962) | 3.57 (1.28) | 0.6540 (0.2527) | 0.3272 (0.2559) |
| **Male** | 41.42 (17.88) | 0.2578 (0.0911) | 3.13 (1.23) | 0.6893 (0.2482) | 0.2950 (0.2497) |

**Table 3. 24-Hour Glucose Dynamics Across Patient Subgroups**. Key CGM-derived statistics stratified by diabetes type, age group, and gender. The CGM sequences analyzed represent the prediction input windows from all test sets. Metrics include glucose variability (standard deviation and coefficient of variation), glycemic dynamics (mean rate of glucose change), and glycemic control indicators: time in range (TIR, 70–180 mg/dL) and time above range (TAR, >180 mg/dL).

To explore the higher prediction errors observed in specific subgroups—particularly females, younger adults, and patients with Type 1 diabetes—we conducted an error analysis based on CGM-derived features. As summarized in Table 3, these groups exhibited greater glucose variability (higher standard deviation and coefficient of variation), faster glucose shifts (higher mean rate of change), and less time spent in the healthy glucose range (lower TIR). For example, females had a standard deviation of 45.18 mg/dL compared to 41.42 mg/dL in males, along with a higher time-above-range (32.7% vs. 29.5%). Similarly, the 18–39 age group showed the highest variability and fastest glucose dynamics. These physiological patterns likely contribute to increased modeling difficulty and reduced prediction accuracy.

## Performance across Prediction Scenarios

We evaluated CGM-LSM's prediction performance across diverse real-world contexts, focusing on two key factors at prediction time: the hour of day and the current glucose level. These factors reflect underlying physiological rhythms and behavioral influences that affect blood glucose dynamics.

Human glucose generation is influenced by lifestyle factors like diet, exercise, and medications. Glucose levels and fluctuations vary by time of day due to their association with human activities like eating and exercise. Given this link, understanding CGM-LSM's performance at different hours is crucial for its safe application. Figure 3A-B shows CGM-LSM's prediction performance at different hours of the day, measured via rMSE over 30-minute, one-hour, and two-hour horizons using the held-out set for T1D and T2D patients. Each line in Figure 4 represents the model's rMSE values across different hours of the day. The model performed consistently well across prediction settings. Clear trends emerged based on prediction horizons and time of day. Shorter horizons (e.g., 30 minutes) had lower rMSE, reflecting better capture of short-term fluctuations. The rMSE score peaked at noon and evening hours. The higher rMSE during these hours likely reflects impacts from diet, exercise, and treatment routines. In contrast, rMSE was notably lower at night. Nocturnal glucose levels, which are less influenced by behavior, allowed CGM-LSM to predict more accurately. These findings demonstrate the importance of temporal factors and human behaviors in glucose prediction.

In addition to time-based variation, prediction performance also differed based on the glucose level at the time of prediction (Figure 3C–D). These plots reveal rMSE across a range of glucose values, providing insight into the model's reliability across glycemic states. Performance was strongest in euglycemic ranges (~100–150 mg/dL), where predictions remained relatively stable. However, rMSE rose sharply in both hypoglycemic (<70 mg/dL) and hyperglycemic (>250 mg/dL) ranges—particularly for longer horizons. This trend likely reflects a combination of physiological volatility and data imbalance. Extreme glucose states are rarer in the training set, leading to undertrained token embeddings in autoregressive prediction. Furthermore, glucose behavior in these regions is more dynamic and less predictable due to compensatory mechanisms (e.g., insulin action, counterregulation, carb intake). These findings emphasize that model performance is not uniform across the glucose spectrum and underscore the need for cautious interpretation in edge cases.

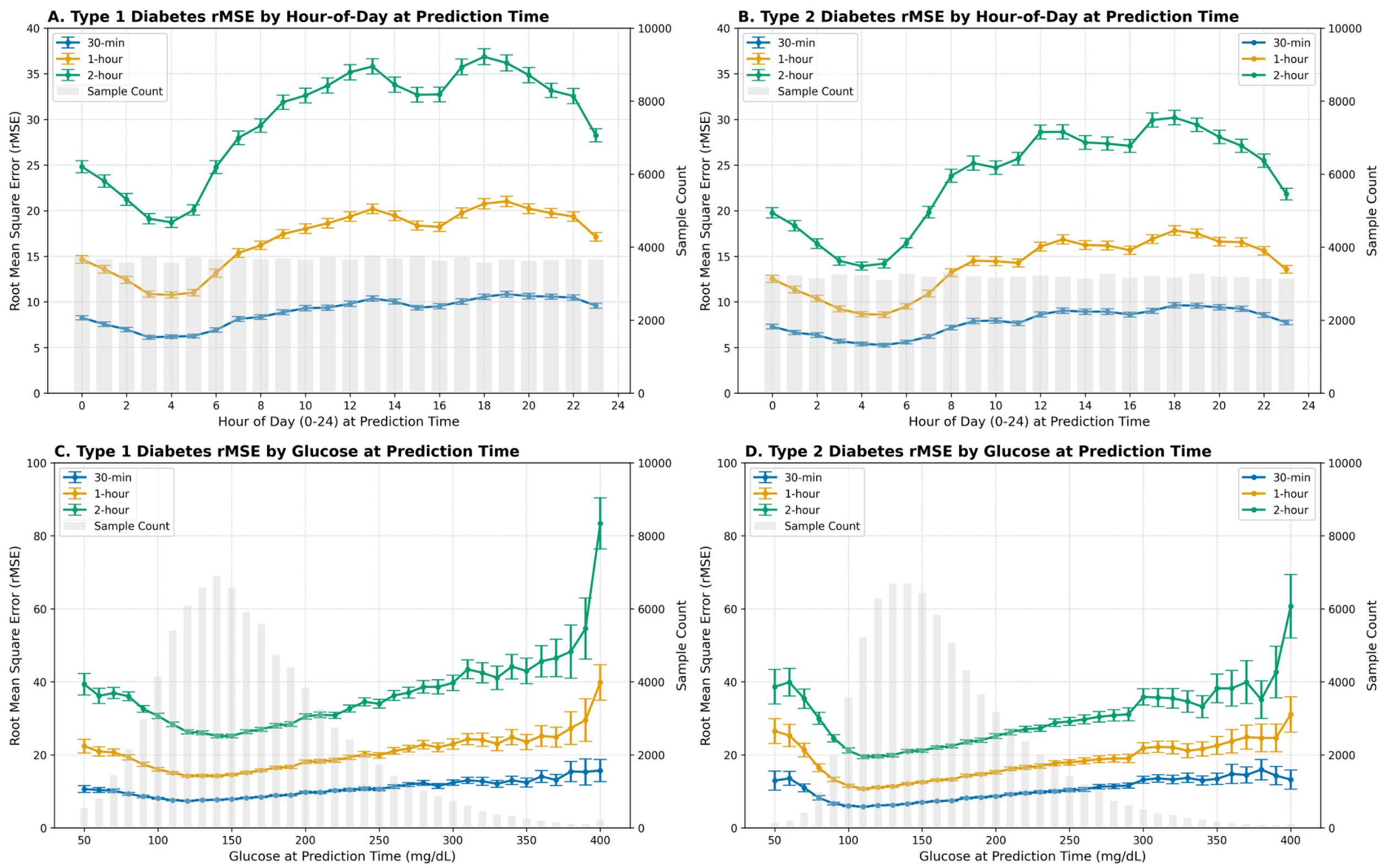

**Figure 4. CGM-LSM Forecast Performance Across Prediction Scenarios. A–B**: Forecast error (rMSE) by hour of prediction for patients with Type 1 (A) and Type 2 (B) Diabetes across 24 hours. **C–D**: Forecast error (rMSE) by glucose level at prediction time for Type 1 (C) and Type 2 (D) Diabetes.

# Discussion

Our study demonstrates that CGM-LSM is a robust, high-performance model for glucose forecasting across a range of real-world scenarios. This is made possible in large part by the richness and diversity of our pretraining dataset, which includes over 15 million CGM records from 592 patients spanning different diabetes types, age, and gender. This diversity ensures that the model is exposed to a wide spectrum of glucose trajectories during pretraining, enabling it to learn generalizable temporal patterns.

Our model is a decoder-only Transformer based on the GPT-2 architecture, designed to predict glucose trajectories by modeling CGM values as categorical tokens in an autoregressive manner. Each instance includes a 26-hour CGM sequence, where the model predicts the next token one step at a time based on preceding values. This architecture is particularly suited for glucose forecasting due to its ability to capture both short- and long-range temporal dependencies via self-attention. Unlike RNNs (e.g., LSTM, GRU), which process sequences sequentially and may suffer from vanishing gradients or long context windows. Several prior works have explored various model architectures for glucose prediction including attention-based RNNs[16], CNNs[15], and ensemble approaches[30] but face limited performance improvement. These approaches suffer from: (1) unsuitable model structures that either struggle with long-range dependencies or lack parallel processing capabilities; and (2) insufficient data leading to underfitting, as highlighted by transfer learning and augmentation attempts in recent studies[17]. The Transformer decoder attends over all prior time steps in parallel—effectively modeling complex physiological patterns such as insulin response or circadian rhythms. Importantly, the autoregressive setup aligns naturally with the forecasting task—the model is pretrained to generate the next glucose token based on past values, which directly mirrors the downstream prediction objective. This allows CGM-LSM to learn temporal patterns in a self-supervised fashion.

A key design decision in CGM-LSM is the treatment of glucose values as categorical tokens, learned through a Transformer decoder in an autoregressive fashion. This token-based design accounts for the nonlinear and asymmetric physiological meaning of glucose values. Quantitatively, CGM-LSM outperforms prior methods by a substantial margin. On the OhioT1DM dataset, CGM-LSM reduced rMSE by 48.51% at the one-hour horizon compared to widely used baselines. These gains were consistent across both short-term (30 minutes) and long-term (2 hours) prediction horizons, demonstrating CGM-LSM's ability to capture both immediate fluctuations and longer-range glucose dynamics. Moreover, on the Welldoc dataset, CGM-LSM achieved strong prediction performance in challenging generalization settings, including temporal splits (predicting future periods) and held-out patients. These findings imply that CGM-LSM can serve as a more reliable forecasting tool in dynamic and diverse clinical environments, where patient profiles and care contexts vary over time. The model also demonstrated demographic robustness, performing consistently across patient subgroups defined by age, gender, and diabetes type. This performance evaluation is essential for ensuring fair access and effectiveness across heterogeneous populations, and highlights the model's suitability for broad deployment in real-world diabetes management[16]. We believe CGM-LSM's superior prediction accuracy and robustness are driven by its pretraining on the large-scale CGM data, which allows the model to learn glucose generation mechanisms from sensor curves. Its success indicates the value of applying foundation models to large-scale sensor data for improved prediction.

At the same time, we found subgroups with higher prediction errors (e.g., T1D, younger, or female patients). These differences suggest that certain populations may present greater physiological complexity or variability, potentially requiring more tailored modeling strategies. Motivated by these disparities, we further examined the model's performance across different application scenarios—including time of day and glucose level at time of prediction. Accuracy tended to decline during daytime hours, particularly around meals, and at glucose extremes (e.g., hypoglycemia and hyperglycemia). These are periods when human behavior (e.g., eating, exercising) or physiological instability likely introduce higher variability and fewer representative training examples. These findings are important for two reasons. First, they demonstrate the value of subgroup-aware and context-aware evaluations, extending beyond aggregate accuracy to assess where model performance may falter. Second, these results offer actionable guidance for future foundation model design. On one hand, they motivate personalization efforts, such as subgroup-specific fine-tuning or adaptive calibration. On the other hand, they highlight a need to incorporate human behavior and temporal rhythms more explicitly into model architectures, possibly through auxiliary inputs or multimodal pretraining. While large-scale pretraining enables CGM-LSM to generalize across many settings, these remaining limitations emphasize that scale alone is not sufficient. Continued progress will require models that are not only data-rich but behavior-aware and fairness-conscious, capable of adapting to the diversity inherent in real-world clinical deployment.

Our study has several key limitations that warrant future analysis. First, our approach models glucose values as discrete categorical tokens, enabling the model to learn meaningful representations through token embeddings. However, how a numeric representation strategy might work in a pretrained foundation model remains unexplored and requires further empirical research. Second, many modeling design questions remain unaddressed, such as how model size, model structure, and data scale affect forecasting performance. Investigating whether scaling laws, observed in language models, hold in the CGM context could yield insights into the trade-offs between scale and generalization. Third, we rely solely on CGM sequences without incorporating contextual information such as medication, meals, physical activity, or education events. These behavioral and clinical inputs are known to affect glucose dynamics but are often missing or sparsely recorded. Integrating such data streams into foundation models poses both challenges in both modeling and data availability. Fourth, we did not explore patient-specific fine-tuning or adaptation. While CGM-LSM generalizes well across populations, individual responses to interventions or lifestyle changes are highly personalized. Future work should examine methods for fine-tuning with a few days of CGM to support truly personalized diabetes management. Together, these limitations highlight both the promise and the complexity of building general-purpose foundation models for physiological time series. Addressing them will require innovations in representation learning, model architecture, data integration, and adaptive training strategies.

In conclusion, this study utilizes CGM sensor data to pretrain a large sensor model to improve near-future glucose prediction performance. This paper expands the AI for diabetes management from long-term risk level prediction to near future prediction to enable the patient to action to control the glucose in range. This glimpse to the future will empower the patients more about learning their near-future glucose to improve the diabetes self-management and receive the quick feedback.

# Methods

## Datasets

The protocol was approved by the Johns Hopkins School of Medicine Institutional Review Board (IRB00447704). For this study, we used two datasets for our analysis: CGM data collected from Welldoc, Inc. for pretraining and the OhioT1DM data for benchmarking.

**Welldoc Data**: This dataset, used for pretraining the CGM-LSM model, initially contained 21,215,912 CGM records from 617 patients. The data were de-identified, with only patient age, gender, and diabetes type available for analysis.

To ensure compliance with de-identification requirements, we applied a random-day shift method to the datetime values associated with each patient, consistent with the approach utilized in the MIMIC-III dataset[31]. This technique enhances data security by anonymizing the temporal aspects of the dataset while preserving the relative patterns and temporal coherence of the datetime information for each individual. By maintaining the internal structure of temporal data, this method facilitates meaningful analysis without compromising patient confidentiality.

We constructed instances to align with the context where for a given patient and an instance datetime, the model predicted future health outcomes based on available patient information. Thus, each instance was associated with a patient and a specific instance datetime. Instance datetimes were derived from the entry times recorded in the CGM system. In the application scenarios, when new glucose data are recorded from the CGM system, our model is triggered to observe patient information and then predict glucose levels for the next two hours.

To ensure that each instance had sufficient glucose readings for effective learning by the model, we applied specific filtration criteria to retain only valid instances from the raw dataset. First, we assessed the adequacy of input data based on the number of CGM records available prior to the instance datetime. Specifically, we required that a patient have 288 CGM readings recorded in the 24 hours preceding the instance datetime to be included. Additionally, to mitigate the influence of consistent measurement errors that might skew the data, we calculated the mode rate of the CGM sequence. Any instance with a mode rate exceeding 40% was excluded to avoid potential noise in the dataset.

Second, for an instance to be considered valid, we imposed the restriction that it contains reliable “output” data for prediction. This required the data to include a complete record of the subsequent two hours of CGM data (24 CGM records) following the instance datetime. We imposed the requirement of a mode rate below 40% for these two-hour CGM records as well.

We chose a 24-hour window prior to the instance datetime to encapsulate daily patterns and provide a comprehensive view of a patient’s typical day. The subsequent two-hour window was selected because it is deemed clinically significant. Compared to shorter spans such as 30 minutes or one hour, a two-hour prediction window offers a balance of enabling accurate short-term forecasts and avoiding the increased uncertainty associated with longer prediction horizons, such as three hours, which may introduce unreliability due to numerous unobserved variables.

After filtering the data as described above, we retained 15,961,183 valid CGM records from 592 individuals. Of these patients, 291 have type 1 diabetes (T1D) and 301 have type 2 diabetes (T2D). The participants are 52.2% male (309 individuals) and 47.8% female (283 individuals). The dataset includes time-stamped glucose records that were utilized for model training. Compared with previous benchmark data, our novel Welldoc dataset is substantially larger, with CGM data from diverse patient types.

**OhioT1DM Data**: The OhioT1DM CGM dataset[19] is a publicly available collection of continuous glucose monitoring (CGM) data from patients with type 1 diabetes. The dataset was divided into a training and a test set. For comparison with baseline models from previous studies, we focused on the test dataset, which included 166,532 CGM records from 12 patients with time-stamped glucose values. After applying the inclusion criteria used for Welldoc data, we retained 58,414 valid instances from 12 T1D patients. This dataset did not include patient demographic information.

## Training and evaluation datasets

To ensure robust evaluation of model performance in terms of accuracy and robustness, we adopted a stringent methodology for splitting data. For the Welldoc dataset, a held-out test set, comprising 10% of the patients, was excluded from the training process and reserved exclusively for evaluation. We also used the OhioT1DM data as a second held-out test set.

The rest dataset, included 90% of the patients, was processed further. We ordered each patient's CGM recorded glucose values chronologically by instance datetime. The latest 10% of these instances were designated as the "temporal" test set, reserved for temporal validation. The remaining 90% of instances were randomly divided, with 10% forming the "internal" test set and 80% designated for training. This approach allowed the model to be tested against both time periods as well as against patients it had not previously encountered, thereby enhancing the evaluation of its predictive capabilities.

To prevent overfitting and to improve computational efficiency, we applied a random 10% down-sampling strategy to each dataset. This down-sampling was strategically chosen based on the proximity of instance datetimes within instances. Closely timed instances often have substantial overlap in their CGM inputs and outputs, potentially skewing the model's learning with redundant information. For example, if a patient had two instances with instance datetimes of 11:00 am and 11:30 am on the same day, their CGM input data and CGM output data would overlap substantially. This down-sampling thus ensured a sparser distribution of instance times, optimizing training efficiency. All random splits were performed using a fixed random seed of 42 to ensure reproducibility.

Additionally, since each instance was associated with a specific patient and datetime, we were able to subdivide the data further into more granular datasets based on available patient characteristics of age, gender, and diabetes type, and the temporal feature of time of day. To illustrate, using the held-out set as an example, we generated specific test sets based on patient characteristics, such as "held-out T1D," "held-out T1D female," or "held-out T1D 18-30 years old." Similarly, we created test sets based on datetime features, such as "held-out T1D 2pm," which would include instances from T1D patients in the held-out group with instance datetimes at 2pm.

## Problem definition for pretraining

We represented a sequence of 312 CGM values with timestamps as $(s_1, s_2, s_3, \dots s_n)$ for a 26-hour period (24 hours before the focal datetime and 2 hours after). This sequence has a natural sequential ordering, similar to natural language. Following the logic of a language model, the large sensor model is trained to use the autoregression method to predict the next glucose value from the CGM sensor data sequence.

Mathematically, this can be described as maximizing the likelihood of a glucose value $s_i$ given the preceding glucose values $s_1, \dots, s_{i-1}$. The objective function for training the CGM-LSM model is formalized as follows:

$$L(\theta) = \sum_{i=1}^{n} \log p_\theta \left(s_i | s_1, \dots, s_{i-1}\right)$$

where θ represents the parameters of the model, $n$ is the length of the input sequence, and $p_\theta(s_i | s_1, \dots, s_{i-1})$ is the probability of the glucose value $s_i$ conditioned on the prior sequence of tokens, as computed by the model. This training paradigm encourages the development of internal representations that capture glucose generation mechanism dependencies between elements in the input sequence, enabling the model to generate coherent and contextually relevant glucose values in future glucose prediction tasks. By repeatedly adjusting $\theta$ to maximize $L(\theta)$, the CGM-LSM model learns to anticipate subsequent glucose value effectively. In our CGM-LSM model, we adopted the popular transformer decoder-only structure to model the above task, and we conducted our pre-training task based on the CGM data.

## Problem definition for prediction/generation

After the CGM-LSM was trained, it could be used to generate new glucose sequences. The generation process of CGM-LSM is similar to that of the GPT models. It involves iteratively sampling a new generated glucose value (token) from a probability distribution of possible next glucose values (tokens), conditioned on a sequence of input glucose values (tokens). This is mathematically represented as:

$$\mathrm{p}(\mathrm{s}_{\mathrm{i}+1} | \mathrm{s}_1, \dots, \mathrm{s}_\mathrm{i}; \theta) = \mathrm{softmax}(\mathrm{h}_\mathrm{i} \mathrm{W})$$

where $\theta$ denotes the model parameters, $h_i$ is the hidden state derived from transformer blocks, and **W** is the output projection matrix. The generation starts with an initial sequence, i.e., the preceding 24-hour CGM data sequence. Each timestep's hidden glucose state is computed and used to generate a probability distribution over the vocabulary from which the glucose value in the next time stamp is sampled. This can be done using random sampling or techniques like top-k or top-p sampling to balance creativity and coherence. For our application, we used the greedy method, which samples the glucose value with the highest probability. This predicted glucose value is then treated as the real value and appended to the original sequence for the next glucose value prediction.

The process repeats until the maximum length, i.e., 24 glucose values for the next two hours, is reached. Finally, the generated 2-hour glucose values are compared with the real glucose values to measure the model's prediction performance.

## CGM-LSM model structure

**Input Representation.** The tokens used in CGM-LSM for glucose prediction were the glucose values themselves. Various methods exist for converting numerical values into categorical tokens. For simplicity and without losing generality, we directly used the glucose readings as tokens; for instance, a glucose level of 153 becomes the token "153". In total, we had 400 glucose value tokens in the vocabulary because the maximum value of glucose measured by the CGM device is 400. In the pretraining tasks involving next glucose value prediction, the model was expected to learn the semantic meanings of the embeddings for these glucose value tokens.

**Model Structure.** CGM-LSM adopted the transformer decoder-only structure[27], which is widely used in GPT model series[20,32]. In contrast to traditional recurrent neural networks[23], this approach relies entirely on self-attention mechanisms to generate sequences. Such an architecture facilitates more parallelizable computations and effectively captures long-range dependencies in data.

In a decoder-only transformer, the model is composed exclusively of a stack of decoder blocks that process the input sequence to generate one glucose value token at a time. Each decoder block in the transformer comprises two main components: a multi-head self-attention mechanism and a position-wise fully connected feed-forward network.

The input to each decoder layer is initially transformed into three vectors—query (Q), key (K), and value (V) vectors—through linear projection of the embeddings of the input tokens. The self-attention mechanism in the decoder is mathematically represented as:

$$\mathrm{Attention}(\mathrm{Q}, \mathrm{K}, \mathrm{V}) = \mathrm{softmax}\left(\frac{\mathrm{QK}^{\mathrm{T}}}{\sqrt{\mathrm{d_k}}}\right)\mathrm{V},$$

where $d_k$ is the dimension of the key vectors. This mechanism allows each position in the decoder to attend to all positions up to and including that position in the previous layer. This is crucial for autoregressive models where the prediction for the next token can only depend on the known previous tokens, thereby preserving the causality in the sequence generation.

Normalization and residual connections are employed around each of these sublayers (self-attention and feed-forward networks), which help in stabilizing the learning process. Mathematically, the output of each sublayer can be described as:

$$\mathrm{LayerNorm}\big(\mathrm{x} + \mathrm{Sublayer}(\mathrm{x})\big),$$

where Sublayer(x) is the function implemented by the sublayer itself, either self-attention or a feed-forward network, and $x$ is the input to the sublayer.

Finally, after passing through the series of decoder blocks, the output is projected onto a vocabulary-sized vector using a linear transformation followed by a softmax layer to produce a probability distribution over possible next tokens:

$$\mathrm{p}(s_{i+1}|\mathrm{s}_1, \dots, \mathrm{s}_i) = \mathrm{softmax}(\mathrm{h_i W}),$$

where $\mathrm{h_i}$ is the final layer's output at position i and $W$ is the weight matrix. The glucose value corresponding to the highest probability is then selected as the next token in the sequence. The decoder-only transformer architecture leverages the power of self-attention to efficiently process sequences in a manner that scales favorably with sequence length and allows for highly parallelizable implementation.

**Loss Function.** Because we treated a glucose value as a unique token, we used cross-entropy to calculate the loss for the next glucose value prediction task. In the pretraining of the CGM-LSM model, the softmax function is instrumental in transforming the logits—raw output scores for each glucose value in the vocabulary—into a probability distribution over all potential next glucose values. This probabilistic framework is crucial for defining the cross-entropy loss, which quantifies the difference between the predicted probabilities and the actual distribution of the next glucose value. Specifically, given the logits $\mathbf{z_t}$, for possible next glucose values at position $t + 1$, the softmax function calculates the probability of each token $x_t$ being the next token in the sequence as follows:

$$\mathrm{p_\theta}(\mathrm{s_{t+1}}|\mathrm{s_1}, \dots, \mathrm{s_t}) = \frac{\exp(\mathrm{z_{t,s_t}})}{\sum_{\mathrm{k=1}}^{\mathrm{V}} \exp(z_{t,k})}$$

where $V$ is the vocabulary size and $z_{t,k}$ represents the logit corresponding to the $k$-th vocabulary token. The cross-entropy loss for predicting a single token is thus:

$$\mathrm{Loss} = -\log \mathrm{p_\theta}\,(\mathrm{s_{t+1}}|\mathrm{s_1}, \dots, \mathrm{s_t})$$

This loss essentially measures the model's effectiveness in predicting the actual next token $\mathrm{s}_{t+1}$ using the probabilities output by the softmax function. By minimizing the negative log-likelihood, or equivalently, the cross-entropy loss across all tokens in the training dataset, the GPT model learns to generate accurate and coherent text. The total loss for a batch of sequences, which aggregates the individual token losses, ensures comprehensive learning across diverse textual contexts:

$$\text{Total Loss} = -\sum_{\mathrm{i=1}}^{\mathrm{N}} \sum_{\mathrm{t=1}}^{\mathrm{T_i}} \log\left(\frac{\exp(z_{i,t-1,s_{i,t}})}{\sum_{k=1}^{V} \exp(z_{i,t-1,k})}\right)$$

Through iterative minimization of this total loss during pretraining, the CGM-LSM model effectively hones its parameter set $\theta$, thereby enhancing its capabilities in language modeling and next-token prediction.

## CGM-LSM implementation details

We utilized the GPT-2 model, as implemented in the Hugging Face library[33], as the foundation for CGM-LSM. The model's token vocabulary contained 17 special tokens and 400 glucose value tokens. For the CGM-LSM structure, the embedding size was 768 the number of multi-heads was 12, and the number of layers was 12.

For model training with the Welldoc training dataset, we set batch size to 256. For the optimizer, we used AdamW[34] with $\beta_1$ to be 0.9 and $\beta_2$ to be 0.999. Grid search was used to obtain the optimal hyperparameters. We set the learning rate to be 0.00005, the layer norm epsilon value as 0.00001, and dropout rate as 0.1. We conducted the training process on a single A100 80GB GPU over ten epochs. Notably, towards the end of the third epoch, we observed an increment in loss on the validation sets. We therefore used the model instance at the end of the third epoch as our final model.

Post-training, the model's capability to generate new sequences was tested using Hugging Face's generation functions on the same type of GPU. We employed the greedy sampling method[35] to predict glucose values over a two-hour horizon, terminating after the 24th token. After that, we calculated evaluation metrics for the model's prediction performance by comparing these predicted glucose sequences against actual glucose data.

## Model evaluation

In this study, we evaluated model performance using four metrics: root Mean Squared Error (rMSE), Mean Absolute Error (MAE), MAE with a tolerance of 10 units (MAEWith10), and Region Accuracy. These metrics capture the precision and reliability of predictive models in continuous glucose monitoring (CGM) and have been widely used in previous studies of CGM glucose prediction[12,18].

**Root Mean Squared Error (rMSE)** is a robust metric used to quantify the average magnitude of the error between predicted values and actual values. It is computed using the formula:

$$\text{rMSE} = \sqrt{\frac{1}{n}\sum_{i=1}^{n}(y_i - \hat{y_i})^2}$$

where $y_i$ are the actual values, $\hat{y_i}$ are the predicted values, and $n$ is the number of instances.

**Mean Absolute Error (MAE)** measures the average magnitude of the errors in prediction without considering their direction. It is simpler and particularly useful when you need to understand errors in the same units as the data. MAE is calculated as:

$$\text{MAE} = \frac{1}{n}\sum_{i=1}^{n}|y_i - \hat{y_i}|$$

**MAE with Tolerance (MAE($\tau$)),** an adaptation of the traditional MAE, measures the proportion of absolute errors that fall within a specified tolerance level, here set at 10 units. This metric is particularly useful in applications where errors within a certain range are considered acceptable. The calculation is:

$$\text{MAE}(\tau) = \frac{1}{n}\sum_{i=1}^{n}\min(\max(0, |y_i - \hat{y_i}| - \tau), 1)$$

**Glucose Region Accuracy** evaluates the accuracy of predictions by categorizing continuous glucose values into predefined regions and comparing these categorical predictions to actual categorizations. The regions are defined as Very Low (<54 mg/dL), Low (54-69 mg/dL), Time-in-Range (TIR) (70-180 mg/dL), High (181-250 mg/dL), and Very High (>250 mg/dL)[36]. The accuracy is then calculated by:

$$\text{Accuracy} = \frac{\text{Number of correct predictions}}{\text{Total number of predictions}}$$

Collectively, these metrics offered us a comprehensive view of the model's performance, each adding a unique dimension to our understanding of model accuracy and reliability in prediction of glucose levels. The use of varied metrics ensured a balanced evaluation that addressed both the magnitude of errors and their clinical significance.

## Baseline model evaluation

To establish baseline performance for glucose prediction on the OhioT1DM dataset, we evaluated a range of time series forecasting models, starting with classical recurrent neural networks and advancing to more sophisticated transformer architectures. Our evaluation began with fundamental recurrent architectures including vanilla RNN[22], followed by the more capable GRU[24] and LSTM[23] networks, which better capture long-term dependencies in glucose time series. We then progressed to advanced transformer-based architectures, including the standard VanillaTransformer[27], the more efficient Informer[37], and the specialized Autoformer[38] designed specifically for time series forecasting. All models were tasked with predicting future glucose values at clinically relevant horizons of 30, 60, and 120 minutes, using input sequences calibrated to capture meaningful patterns in continuous glucose monitoring data. We utilized the Nixtla NeuralForecast library[39] to implement these models efficiently.

For hyperparameter optimization, we implemented an Optuna-based framework that dynamically scales training steps based on dataset size and batch configuration[40]. This approach allows for efficient exploration of the hyperparameter space while maintaining consistent epoch-based training across different model configurations. The framework automatically converts epoch-based training targets to step-based execution, enabling fair comparison between models with varying computational requirements. Each model underwent rigorous optimization with 10-20 trials to identify optimal configurations for hidden dimensions, number of layers, learning rates, and dropout rates. The models were trained on 289 (24-hour) input to predict future 24 (2-hour) values. All experiments were executed on RTX 4090 GPUs.

## Data availability

The OhioT1DM dataset is publicly available at https://webpages.charlotte.edu/rbunescu/data/ohiot1dm/OhioT1DM-dataset.html. The Welldoc dataset is proprietary and can only be accessed upon request with prior approval from Welldoc. This research study was approved by the Johns Hopkins University Institutional Review Board (IRB# IRB00447704). The study is a retrospective analysis, and no patient protected health information (PHI) was used during model training or evaluation.

## Code availability

All the analyses conducted for this study were performed using Python version 3.11.9, the python package PyTorch version 1.9.0, HuggingFace 4.44.2, and datasets version 2.21.0 was used for developing the CGM-LSM learning models. Visualization was obtained using Python packages matplotlib version 3.9 and Pandas version 2.2.2. The code is available in GitHub: https://github.com/JHU-CDHAI/cgmlsm.

## References

1. Einarson TR, Acs A, Ludwig C, Panton UH. Prevalence of cardiovascular disease in type 2 diabetes: a systematic literature review of scientific evidence from across the world in 2007–2017. *Cardiovasc Diabetol*. 2018;17(1):83. doi:10.1186/s12933-018-0728-6

2. Reidy K, Kang HM, Hostetter T, Susztak K. Molecular mechanisms of diabetic kidney disease. *J Clin Invest*. 2014;124(6):2333-2340. doi:10.1172/JCI72271

3. Pop-Busui R, Boulton AJM, Feldman EL, et al. Diabetic Neuropathy: A Position Statement by the American Diabetes Association. *Diabetes Care*. 2017;40(1):136-154. doi:10.2337/dc16-2042

4. Parker ED, Lin J, Mahoney T, et al. Economic Costs of Diabetes in the U.S. in 2022. *Diabetes Care*. 2024;47(1):26-43. doi:10.2337/dci23-0085

5. Dall T, West T, Ritashree Chakrabarti, Reynolds R, Iacobucci W. 2018 Update The Complexities of Physician Supply and Demand: Projections from 2016 to 2030 Final Report Association of American Medical Colleges. Published online 2018. doi:10.13140/RG.2.2.25694.48963

6. Sheng B, Guan Z, Lim LL, et al. Large language models for diabetes care: Potentials and prospects. *Science Bulletin*. 2024;69(5):583-588. doi:10.1016/j.scib.2024.01.004

7. Kolb L. An Effective Model of Diabetes Care and Education: The ADCES7 Self-Care Behaviors™. *The Science of Diabetes Self-Management and Care*. 2021;47(1):30-53. doi:10.1177/0145721720978154

8. Midyett LK. One Size Fits All Versus Individualized Medicine in Type 1 Diabetes Management. *Diabetes Technology & Therapeutics*. 2023;25(S3):S-42. doi:10.1089/dia.2023.0109

9. Klonoff DC. Continuous Glucose Monitoring. *DIABETES CARE*. 2005;28(5).

10. Galindo RJ, Aleppo G, Parkin CG, et al. Increase Access, Reduce Disparities: Recommendations for Modifying Medicaid CGM Coverage Eligibility Criteria. *J Diabetes Sci Technol*. 2024;18(4):974-987. doi:10.1177/19322968221144052

11. Vettoretti M, Cappon G, Facchinetti A, Sparacino G. Advanced Diabetes Management Using Artificial Intelligence and Continuous Glucose Monitoring Sensors. *Sensors (Basel)*. 2020;20(14):3870. doi:10.3390/s20143870

12. Doorn WPTM van, Foreman YD, Schaper NC, et al. Machine learning-based glucose prediction with use of continuous glucose and physical activity monitoring data: The Maastricht Study. *PLOS ONE*. 2021;16(6):e0253125. doi:10.1371/journal.pone.0253125

13. Zecchin C, Facchinetti A, Sparacino G, De Nicolao G, Cobelli C. Neural Network Incorporating Meal Information Improves Accuracy of Short-Time Prediction of Glucose Concentration. *IEEE Transactions on Biomedical Engineering*. 2012;59(6):1550-1560. doi:10.1109/TBME.2012.2188893

14. Li K, Daniels J, Liu C, Herrero P, Georgiou P. Convolutional Recurrent Neural Networks for Glucose Prediction. *IEEE Journal of Biomedical and Health Informatics*. 2020;24(2):603-613. doi:10.1109/JBHI.2019.2908488

15. Li K, Liu C, Zhu T, Herrero P, Georgiou P. GluNet: A Deep Learning Framework for Accurate Glucose Forecasting. *IEEE Journal of Biomedical and Health Informatics*. 2020;24(2):414-423. doi:10.1109/JBHI.2019.2931842

16. Zhu T, Li K, Herrero P, Georgiou P. Personalized Blood Glucose Prediction for Type 1 Diabetes Using Evidential Deep Learning and Meta-Learning. *IEEE Transactions on Biomedical Engineering*. 2023;70(1):193-204. doi:10.1109/TBME.2022.3187703

17. Deng Y, Lu L, Aponte L, et al. Deep transfer learning and data augmentation improve glucose levels prediction in type 2 diabetes patients. *npj Digit Med*. 2021;4(1):1-13. doi:10.1038/s41746-021-00480-x

18. Zhu T, Chen T, Kuangt L, Zeng J, Li K, Georgiou P. Edge-Based Temporal Fusion Transformer for Multi-Horizon Blood Glucose Prediction. In: *2023 IEEE International Symposium on Circuits and Systems (ISCAS)*. ; 2023:1-5. doi:10.1109/ISCAS46773.2023.10181448

19. Marling C, Bunescu R. The OhioT1DM Dataset for Blood Glucose Level Prediction: Update 2020. *CEUR Workshop Proc*. 2020;2675:71-74.

20. Radford A, Wu J, Child R, Luan D, Amodei D, Sutskever I. Language Models are Unsupervised Multitask Learners.

21. Lecun Y, Bottou L, Bengio Y, Haffner P. Gradient-based learning applied to document recognition. *Proceedings of the IEEE*. 1998;86(11):2278-2324. doi:10.1109/5.726791

22. Rumelhart DE, Hinton GE, Williams RJ. Learning representations by back-propagating errors. *Nature*. 1986;323(6088):533-536. doi:10.1038/323533a0

23. Hochreiter S, Schmidhuber J. Long Short-Term Memory. *Neural Computation*. 1997;9(8):1735-1780. doi:10.1162/neco.1997.9.8.1735

24. Cho K, van Merriënboer B, Gulcehre C, et al. Learning Phrase Representations using RNN Encoder–Decoder for Statistical Machine Translation. In: Moschitti A, Pang B, Daelemans W, eds. *Proceedings of the 2014 Conference on Empirical Methods in Natural Language Processing (EMNLP)*. Association for Computational Linguistics; 2014:1724-1734. doi:10.3115/v1/D14-1179

25. Jin M, Wang S, Ma L, et al. Time-LLM: Time Series Forecasting by Reprogramming Large Language Models. Published online January 29, 2024. doi:10.48550/arXiv.2310.01728

26. Ansari AF, Stella L, Turkmen C, et al. Chronos: Learning the Language of Time Series. Published online May 2, 2024. doi:10.48550/arXiv.2403.07815

27. Vaswani A, Shazeer N, Parmar N, et al. Attention is All you Need. In: *Advances in Neural Information Processing Systems*. Vol 30. Curran Associates, Inc.; 2017. Accessed July 19, 2024. https://proceedings.neurips.cc/paper/2017/hash/3f5ee243547dee91fbd053c1c4a845aa-Abstract.html

28. Kalyani RR, Egan JM. Diabetes and Altered Glucose Metabolism with Aging. *Endocrinology and Metabolism Clinics*. 2013;42(2):333-347. doi:10.1016/j.ecl.2013.02.010

29. Tramunt B, Smati S, Grandgeorge N, et al. Sex differences in metabolic regulation and diabetes susceptibility. *Diabetologia*. 2020;63(3):453-461. doi:10.1007/s00125-019-05040-3

30. Zhu T, Uduku C, Li K, Herrero P, Oliver N, Georgiou P. Enhancing self-management in type 1 diabetes with wearables and deep learning. *npj Digit Med*. 2022;5(1):1-11. doi:10.1038/s41746-022-00626-5

31. Johnson AEW, Pollard TJ, Shen L, et al. MIMIC-III, a freely accessible critical care database. *Sci Data*. 2016;3(1):160035. doi:10.1038/sdata.2016.35

32. Brown T, Mann B, Ryder N, et al. Language Models are Few-Shot Learners. In: *Advances in Neural Information Processing Systems*. Vol 33. Curran Associates, Inc.; 2020:1877-1901. Accessed July 19, 2024. https://proceedings.neurips.cc/paper/2020/hash/1457c0d6bfcb4967418bfb8ac142f64a-Abstract.html

33. Wolf T, Debut L, Sanh V, et al. Transformers: State-of-the-Art Natural Language Processing. In: Liu Q, Schlangen D, eds. *Proceedings of the 2020 Conference on Empirical Methods in Natural Language Processing: System Demonstrations*. Association for Computational Linguistics; 2020:38-45. doi:10.18653/v1/2020.emnlp-demos.6

34. Loshchilov I, Hutter F. Decoupled Weight Decay Regularization. Published online January 4, 2019. Accessed July 19, 2024. http://arxiv.org/abs/1711.05101

35. Wiher G, Meister C, Cotterell R. On Decoding Strategies for Neural Text Generators. *Transactions of the Association for Computational Linguistics*. 2022;10:997-1012. doi:10.1162/tacl_a_00502

36. Battelino T, Danne T, Bergenstal RM, et al. Clinical Targets for Continuous Glucose Monitoring Data Interpretation: Recommendations From the International Consensus on Time in Range. *Diabetes Care*. 2019;42(8):1593-1603. doi:10.2337/dci19-0028

37. Zhou H, Zhang S, Peng J, et al. Informer: Beyond Efficient Transformer for Long Sequence Time-Series Forecasting. *Proceedings of the AAAI Conference on Artificial Intelligence*. 2021;35(12):11106-11115. doi:10.1609/aaai.v35i12.17325

38. Wu H, Xu J, Wang J, Long M. Autoformer: decomposition transformers with auto-correlation for long-term series forecasting. In: *Proceedings of the 35th International Conference on Neural Information Processing Systems*. NIPS '21. Curran Associates Inc.; 2021:22419-22430.

39. Nixtla/neuralforecast: Scalable and user friendly neural forecasting algorithms. Accessed July 27, 2025. https://github.com/Nixtla/neuralforecast

40. Akiba T, Sano S, Yanase T, Ohta T, Koyama M. Optuna: A Next-generation Hyperparameter Optimization Framework. In: *Proceedings of the 25th ACM SIGKDD International Conference on Knowledge Discovery & Data Mining*. KDD '19. Association for Computing Machinery; 2019:2623-2631. doi:10.1145/3292500.3330701

# Supplementary Appendix

This appendix accompanies ***A Large Sensor Foundation Model Pretrained on Continuous Glucose Monitoring Data for Diabetes Management*** and been provided by the authors to give readers additional information about this paper.

Junjie Luo, M.S.,[1,2] Abhimanyu Kumbara, M.S.,[3] Mansur Shomali, M.D.,[3] Rui Han, M.S.,[2] Anand Iyer, Ph.D., M.B.A.,[3] Grazia Aleppo, M.D.,[4] Ritu Agarwal, Ph.D., M.B.A.,[2] Gordon Gao, Ph.D., M.B.A.[2]

[1] School of Medicine, Johns Hopkins University
[2] The Center for Digital Health and Artificial Intelligence, Carey Business School, Johns Hopkins University
[3] Welldoc Inc.
[4] Feinberg School of Medicine, Northwestern University

## Table of Contents

# 1. Supplementary Figures and Tables

## 1.1 Figure S1: Data Construction for the Welldoc Dataset

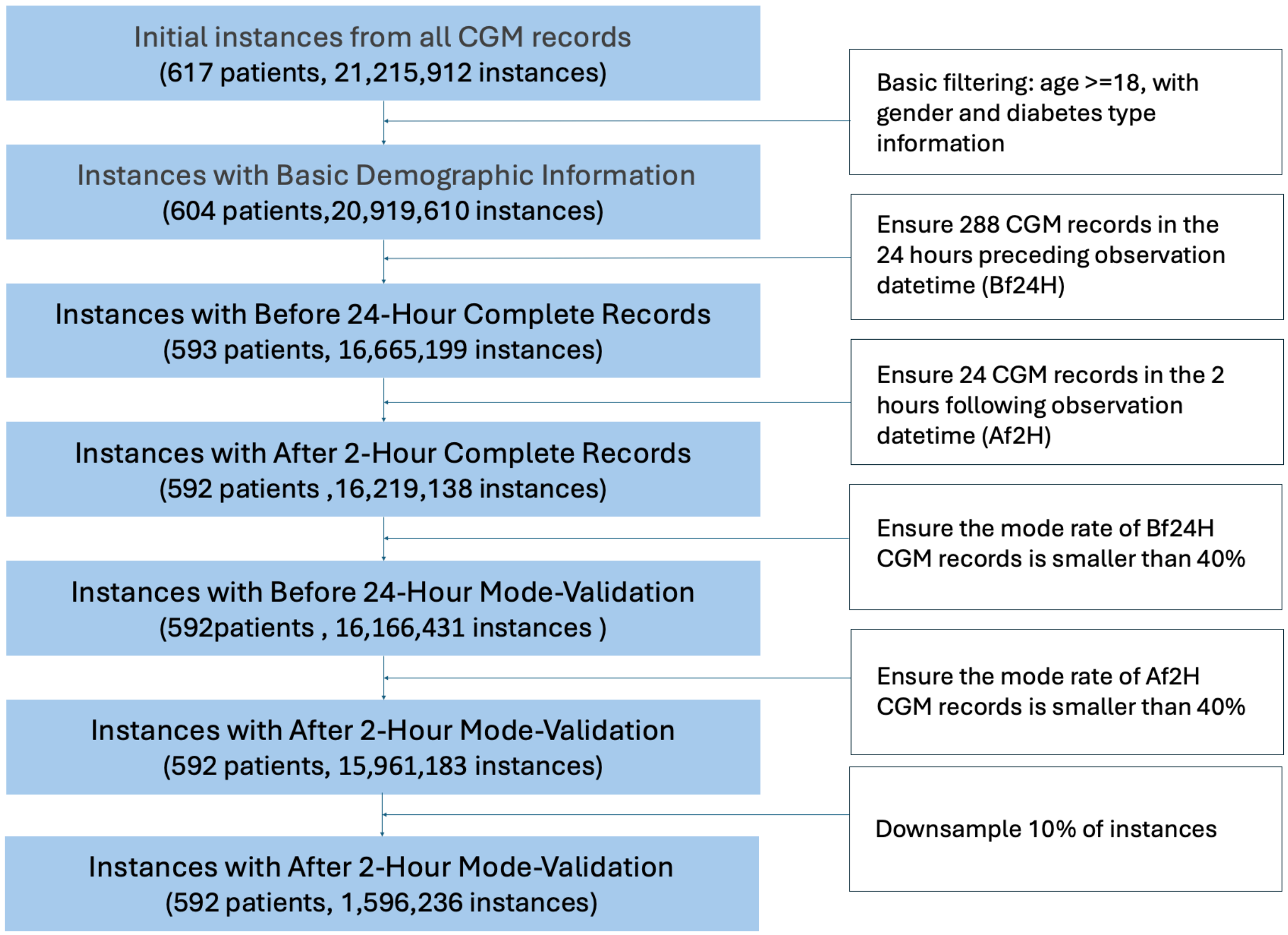

**Figure S1:** The data construction process for the Welldoc Dataset.

## 1.2 Figure S2: Data Construction for the OhioT1DM Dataset

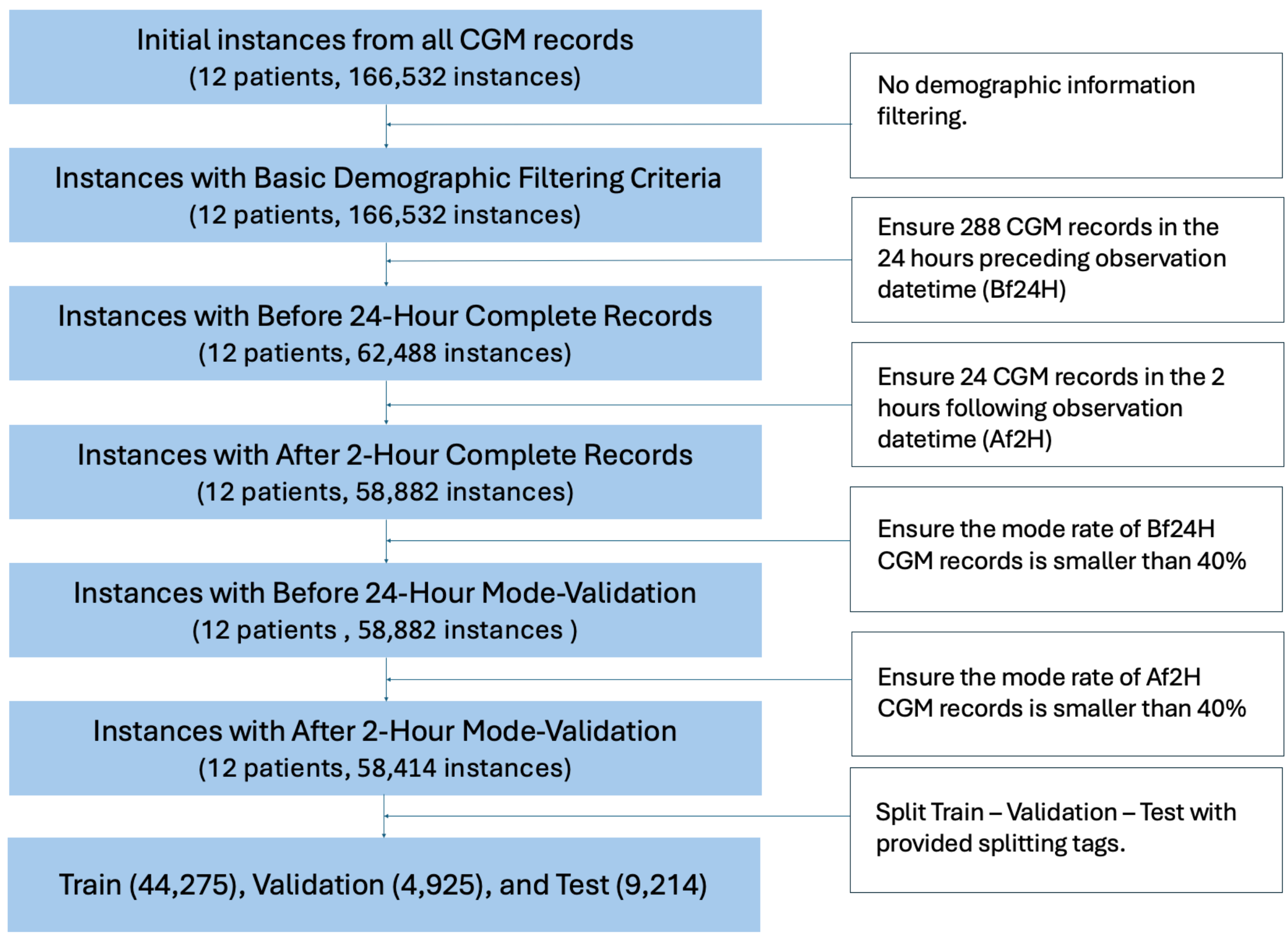

**Figure S2:** The data construction process for the OhioT1DM Dataset.

## 1.3 Table S1: Model Results with rMSE

| Diabetes Type | Evaluation Set | rMSE-30Min | rMSE-1stH | rMSE-2H |
|---|---|---|---|---|
| Type 1 Diabetes | OhioT1DM | 9.358 (0.684) | 15.64 (1.078) | 26.296 (1.664) |
| Type 1 Diabetes | Internal Test | 8.403 (0.066) | 16.049 (0.118) | 28.277 (0.188) |
| | Temporal Test | 9.155 (0.068) | 17.013 (0.118) | 29.426 (0.184) |
| | Hold-Out Test | 8.926 (0.056) | 16.905 (0.101) | 29.812 (0.16) |
| Type 2 Diabetes | Internal Test | 7.441 (0.055) | 13.418 (0.094) | 22.649 (0.147) |
| | Temporal Test | 8.025 (0.058) | 14.073 (0.095) | 23.216 (0.143) |
| | Hold-Out Test | 7.772 (0.055) | 13.877 (0.091) | 23.494 (0.143) |

Table S1: Model results with the rMSE metric.

## 1.4 Table S2: Model Results with MAE

| Diabetes Type | Evaluation Set | MAE-30Min | MAE-1stH | MAE-2H |
|---|---|---|---|---|
| Type 1 Diabetes | OhioT1DM | 8.119 (0.608) | 13.408 (0.957) | 22.178 (1.449) |
| Type 1 Diabetes | Internal Test | 7.001 (0.056) | 13.295 (0.099) | 23.461 (0.16) |
| | Temporal Test | 7.691 (0.058) | 14.205 (0.101) | 24.552 (0.158) |
| | Hold-Out Test | 7.497 (0.048) | 14.099 (0.085) | 24.889 (0.138) |
| Type 2 Diabetes | Internal Test | 6.246 (0.047) | 11.19 (0.08) | 18.876 (0.125) |
| | Temporal Test | 6.805 (0.05) | 11.833 (0.081) | 19.472 (0.123) |
| | Hold-Out Test | 6.571 (0.047) | 11.64 (0.078) | 19.678 (0.123) |

Table S2: Model results with the MAE metric.

## 1.5 Table S3: Model Results with MAE (10)

| Diabetes Type | Evaluation Set | MAE(10)-30Min | MAE(10)-1stH | MAE(10)-2H |
|---|---|---|---|---|
| Type 1 Diabetes | OhioT1DM | 0.747 (0.024) | 0.587 (0.026) | 0.42 (0.022) |
| Type 1 Diabetes | Internal Test | 0.7921 (0.0021) | 0.6132 (0.0024) | 0.4383 (0.0022) |
| | Temporal Test | 0.7681 (0.0021) | 0.5912 (0.0023) | 0.4229 (0.0021) |
| | Hold-Out Test | 0.7717 (0.0018) | 0.5879 (0.002) | 0.4125 (0.0018) |
| Type 2 Diabetes | Internal Test | 0.8179 (0.002) | 0.6578 (0.0023) | 0.4927 (0.0022) |
| | Temporal Test | 0.7986 (0.002) | 0.6405 (0.0023) | 0.4803 (0.0022) |
| | Hold-Out Test | 0.8074 (0.0019) | 0.6463 (0.0022) | 0.4801 (0.0021) |

Table S3: Model results with the MAE(10) metric.

## 1.6 Table S4: Model Results with Region-Accuracy

| Diabetes Type | Evaluation Set | RegionAccu-30Min | RegionAccu-1stH | RegionAccu-2H |
|---|---|---|---|---|
| Type 1 Diabetes | OhioT1DM | 0.902 (0.018) | 0.842 (0.021) | 0.763 (0.024) |
| Type 1 Diabetes | Internal Test | 0.9352 (0.0014) | 0.8803 (0.0018) | 0.8016 (0.0021) |
| | Temporal Test | 0.9283 (0.0014) | 0.8724 (0.0017) | 0.7932 (0.0021) |
| | Hold-Out Test | 0.925 (0.0012) | 0.8649 (0.0016) | 0.7785 (0.0018) |
| Type 2 Diabetes | Internal Test | 0.9475 (0.0012) | 0.9054 (0.0016) | 0.8457 (0.0019) |
| | Temporal Test | 0.9435 (0.0012) | 0.9017 (0.0015) | 0.8431 (0.0018) |
| | Hold-Out Test | 0.9414 (0.0012) | 0.8964 (0.0015) | 0.8327 (0.0018) |

Table S4: Model results with the Region-Accuracy metric.

## 1.7 Figure S3: Performance Across Age for Different Prediction Horizons

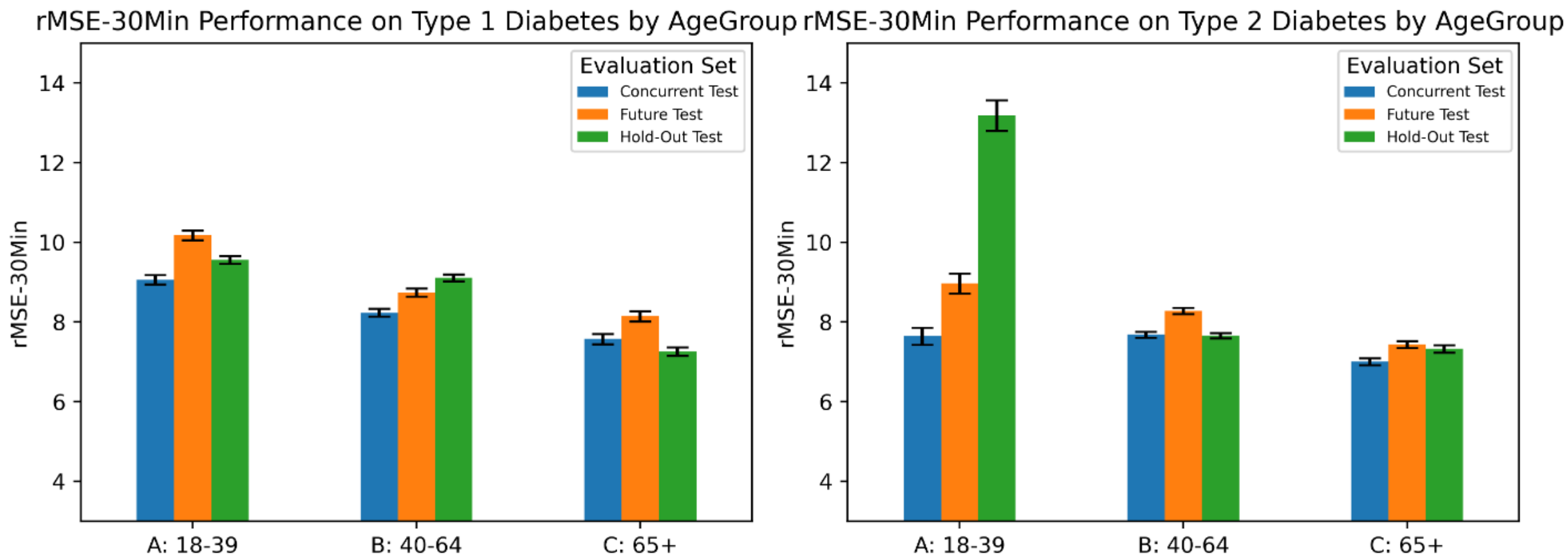

a. Model performance across age groups for 30-minute prediction horizon

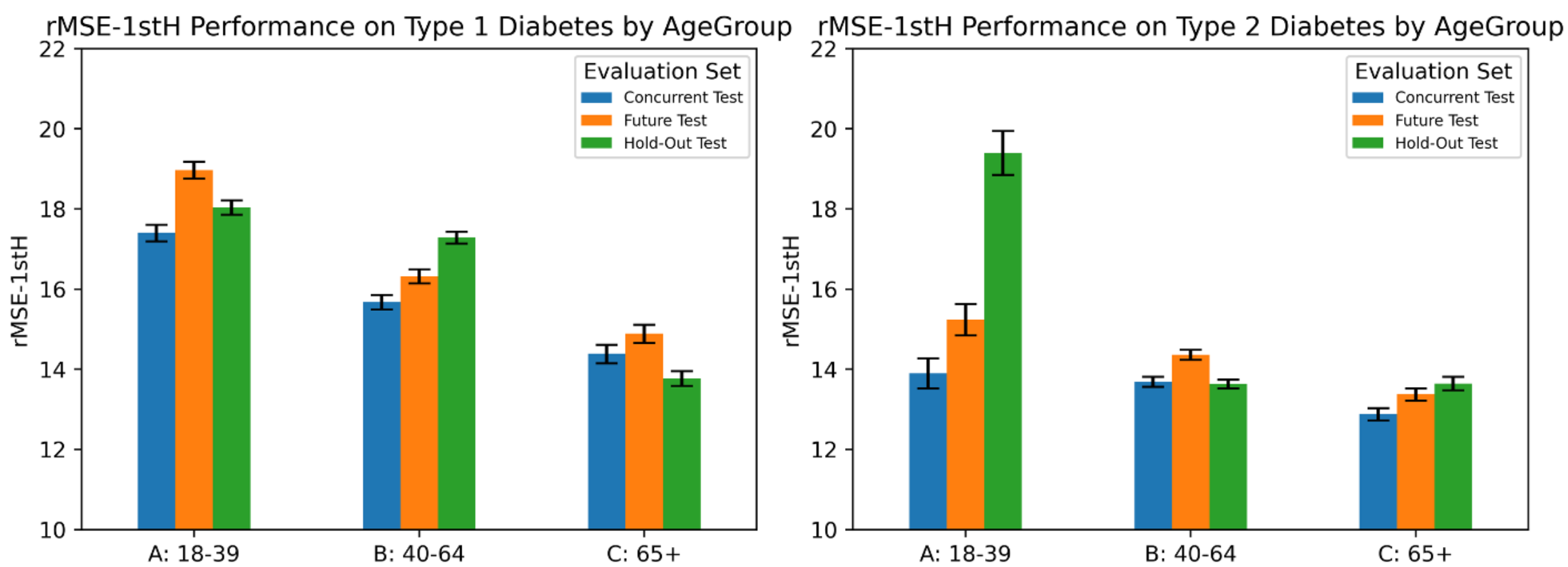

b. Model performance across age groups for one-hour prediction horizon.

**Figure S3:** CGM-LSM's prediction performance in different prediction horizons across age groups.
(a) rMSE-1Hour prediction performance across age groups. (b) rMSE-30Min performance across age groups.

## 1.8 Figure S4: Performance Across Gender for Different Prediction Horizons

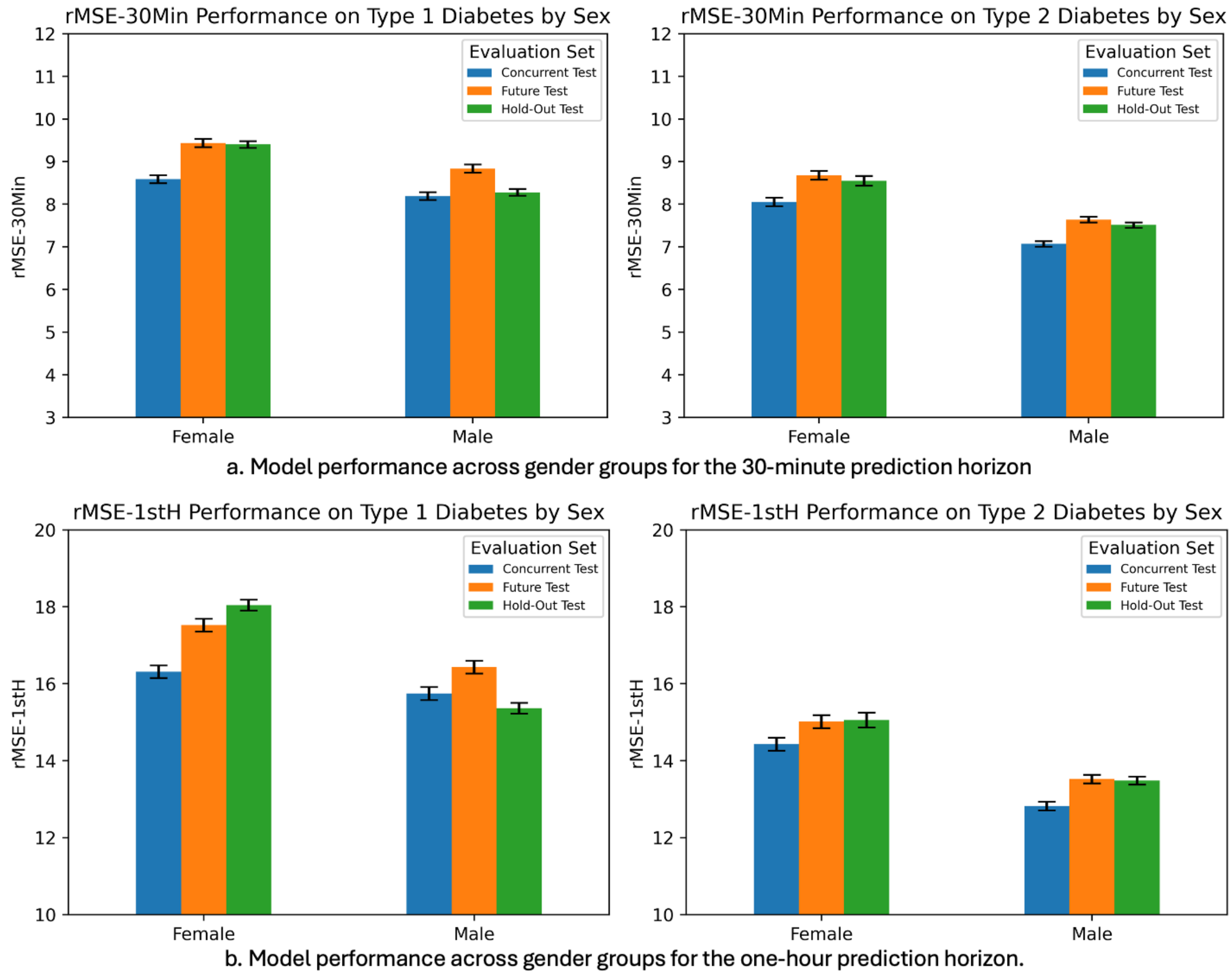

**Figure S4:** CGM-LSM's prediction performances in different prediction horizons across gender groups.
(a) rMSE-1Hour prediction performance across gender groups. (b) rMSE-30Min performance across gender groups.

## 1.9 Figure S5: Performance Across Prediction Hours of the Day on Evaluation Sets.

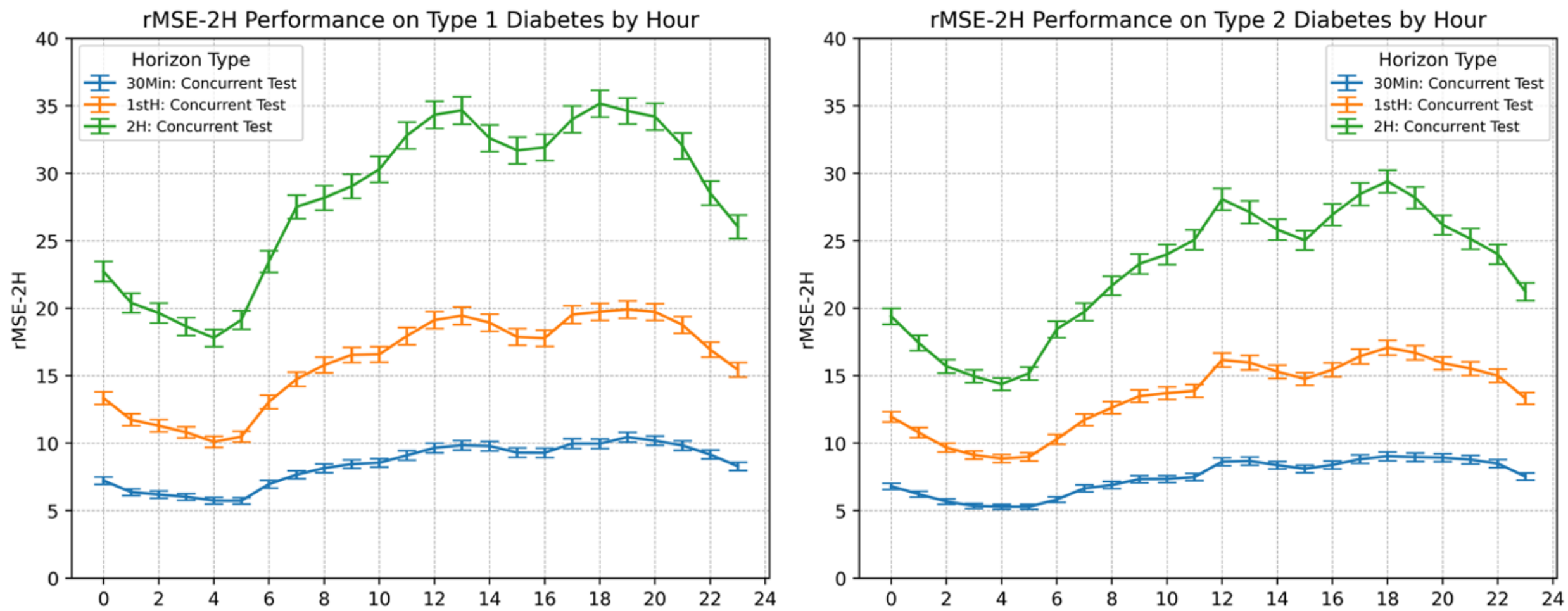

A. Model performance across hour of the day in the concurrent (validation) evaluation set

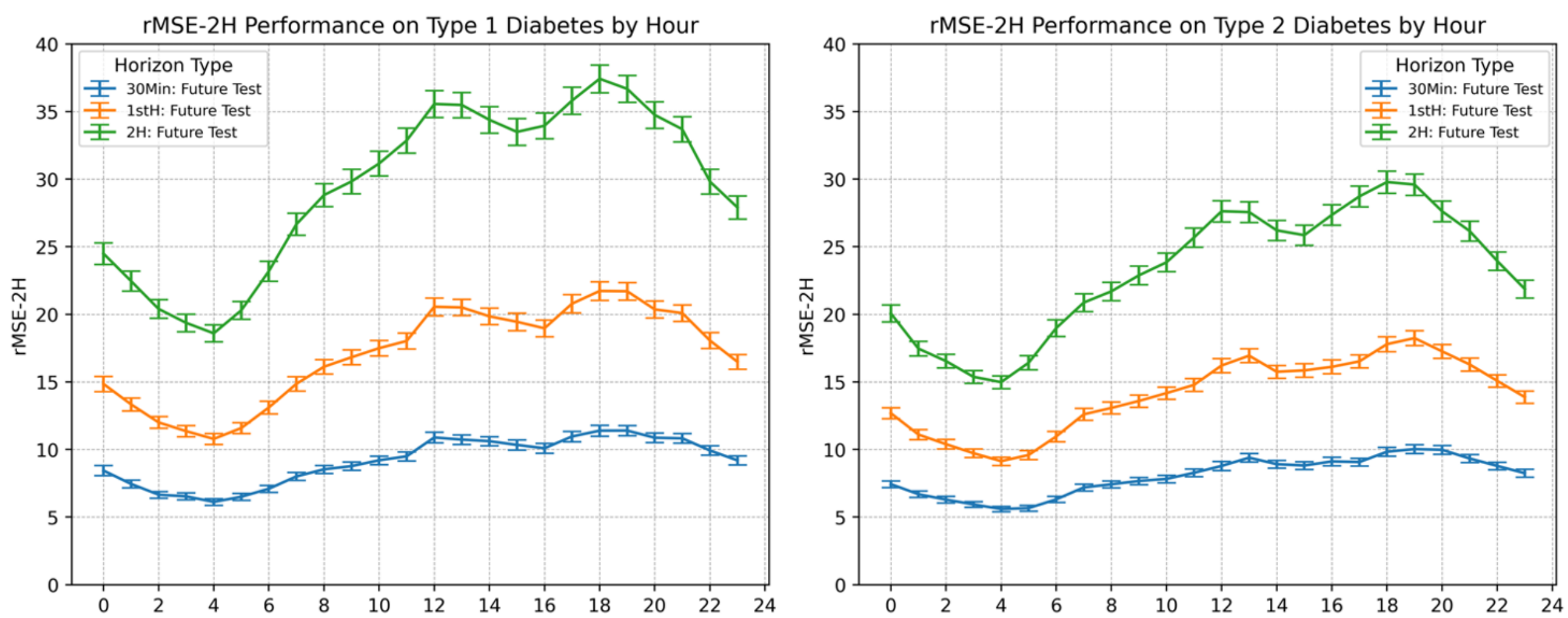

B. Model performance across hour of the day in the future (temporal) evaluation set

**Figure S5:** CGM-LSM's prediction performance across hour of the day. (A). Model performance across hour of the day in the concurrent (validation) evaluation set. (B). Model performance across hour of the day in the future (temporal) evaluation set.

# 1.10 Figure S6: Model Prediction Examples

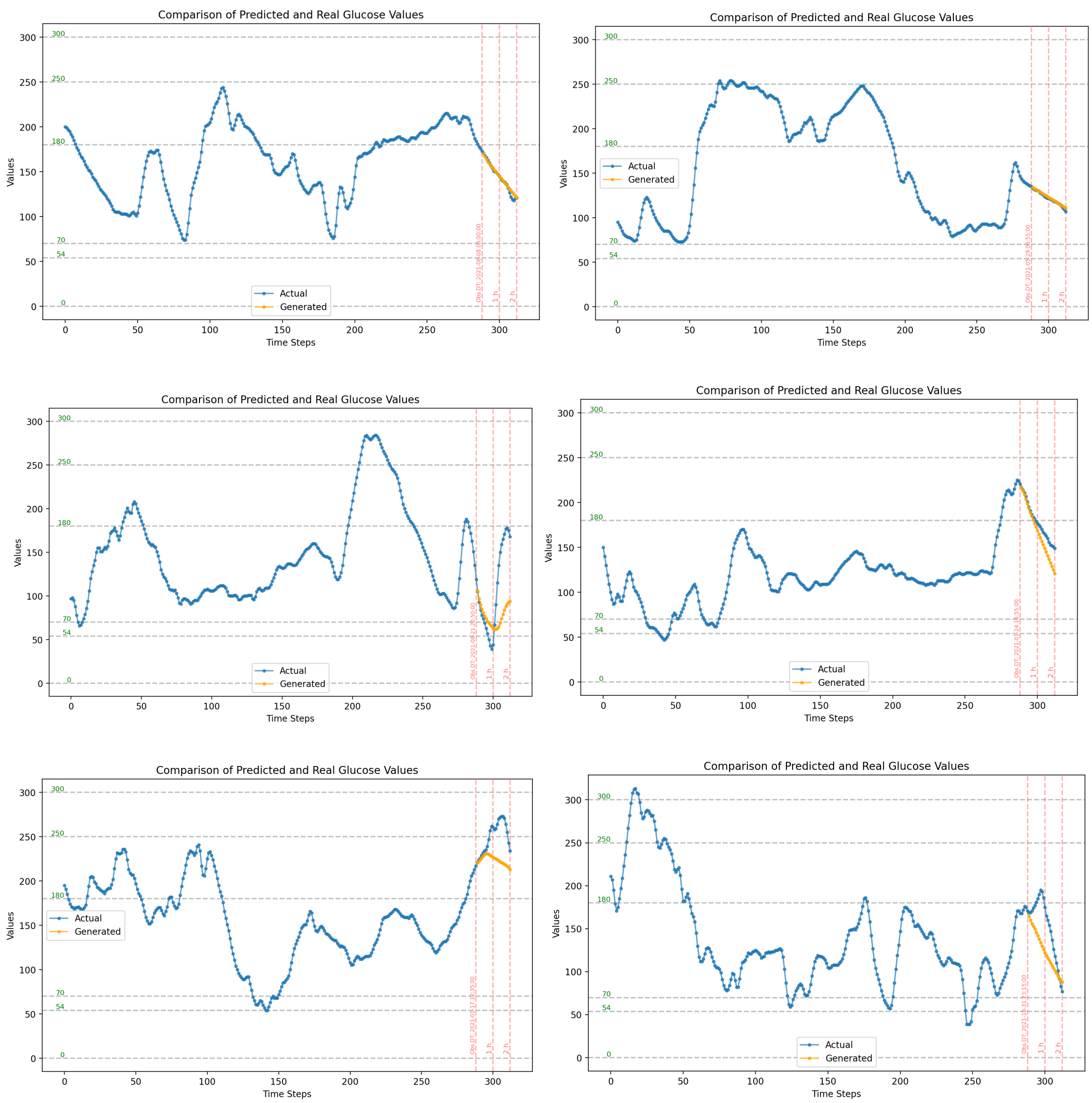

**Figure S6**: Examples of good, moderate, and bad predictions made by the model.